\documentclass[preprints,article,submit,pdftex,moreauthors]{Definitions/mdpi_non} 
\usepackage{amsmath,amssymb,bm,slashed}

\newcommand{\beq}{\begin{align}}
\newcommand{\eeq}{\end{align}}
\newcommand{\sbeq}{\begin{subalign}}
\newcommand{\seeq}{\end{subalign}}

\firstpage{1} 
\pubvolume{1}
\issuenum{1}
\articlenumber{0}
\pubyear{2025}
\copyrightyear{2025}
\datereceived{ } 
\daterevised{ } 
\dateaccepted{ } 
\datepublished{ } 
\hreflink{https://doi.org/} 

\Title{Soft mode dynamics associated with QCD critical point and color superconductivity --- pseudogap, 
anomalous dilepton production and electric conductivity\footnote{Report numbers: YITP-26-39;\,J-PARC-TH-0334}}

\TitleCitation{Soft mode dynamics associated with QCD critical point and color superconductivity }

\Author{Masakiyo Kitazawa$^{1}$*\orcidA{}
and Teiji Kunihiro $^{2}$\orcidB{}}

\address{%
$^{1}$ \quad
Yukawa Institute for Theoretical Physics, 
Kyoto University; J-PARC Branch, KEK Theory Center, Institute of Particle and Nuclear Studies, KEK, Tokai, Ibaraki 319-1106, Japan; kitazawa@yukawa.kyoto-u.ac.jp\\
$^{2}$ \quad Yukawa Institute for Theoretical Physics, 
Kyoto University; kunihiro@yukawa.kyoto-u.ac.jp}

\corres{Correspondence: kitazawa@yukawa.kyoto-u.ac.jp}

\abstract{
We give a systematic account of
the soft mode dynamics of QCD critical point and the two-flavor 
color-superconductivity based on the 2-flavor Nambu--Jona-Lasinio model, 
and investigate their effects on  electromagnetic observables
in relativistic heavy-ion collisions (HIC).
We first demonstrate that the collective excitations coupled  to the 
fluctuations of the respective order parameters are the soft modes associated with the 
phase transitions, in the sense that they 
acquire a prominent spectral strength in 
the low-energy and low-momentum region near the phase transitions,
and the peak energy goes down, i.e., gets softened, and eventually vanishes at the critical point.
It is shown that the diquark soft mode of the 2SC gives rise to the pseudogap, i.e., 
a depression in the density of states of the quark
spectra around the Fermi surface above but in the vicinity of 
the critical temperature.
Then, exploiting the ideas that were developed 
in condensed matter physics for describing the `para-conductivity' 
in the normal phase of metal superconductors, we show that 
the soft modes cause an anomalous enhancement of  electric conductivity and 
the dilepton production rate, and discuss their relevance to HIC.
 }


\begin{document}


\section{Introduction}
\label{sec:Introduction}

One of the central problems in modern physics
is to reveal the properties of hot and dense matter
realized in the cores of compact stars and in the early universe.
Since the extremely hot and dense matter should be 
described in terms of quarks and gluons
 governed by quantum chromodynamics (QCD),
such an effort can be tantamount to developing condensed matter physics in terms
of QCD or {\em QCD-condensed matter physics}.
Accumulated  theoretical works based on the lattice QCD simulations and 
low-energy effective theories of QCD show that various types of  phase transitions may occur 
in  such a matter~\cite{Fukushima:2010bq}, which
includes the appearance of the color superconductivity (CSC)~\cite{Alford:2007xm},
and the existence of a critical point called 
the QCD critical point (QCD-CP)~\cite{Asakawa:1989bq,Barducci:1989wi} at which
a first-order phase-transition line in the higher-density
and lower-temperature region ends and the phase transition turns second order with
 nonzero current quark masses\footnote{ 
In fact, the phase structure around the QCD-CP could be more
complicated and might have multiple critical points~\cite{Kitazawa:2002jop}
 due to the vector interaction~\cite{Kunihiro:1991qu}, 
the Kobayashi-Maskawa-'t Hooft (KMT) anomaly term,
and mismatched Fermi spheres of the respective pairing particles.}. 

Terrestrial experiments using high-energy 
heavy-ion collisions (HIC)~\cite{Yagi:2005yb} have been utilized to 
probe such a hot and dense matter.
The beam-energy scan experiments~\cite{Bzdak:2019pkr,STAR:2021iop}
have been carried out to explore dense matter at relatively low temperatures, and next-generation experimental programs~\cite{Galatyuk:2019lcf} are being
developed or planned in several countries. 
Since the fluctuation of the order parameter 
 grows in a divergent way as  the system approaches 
the critical point, 
observables that are coupled to such a fluctuation
should be of primary interest for probing
the QCD phase transitions in HIC.
 
In the case of the QCD-CP~\cite{Stephanov:1998dy}, 
since it comes to exist when an explicit chiral symmetry breaking is present 
at finite baryon density,
 the order parameter of it is a linear
combination of the baryon density and the chiral scalar condensate due to
the violation of the charge-conjugation symmetry, as is familiar with 
Walecka's $\sigma$-$\omega$ model~\cite{Walecka:1974qa,Matsui:1981ag}; see Appendix B of the latter paper.
People have been interested in extracting fluctuations or  
second- and higher-order cumulants of fluctuations of
baryon number densities
through the event-by-event analyses~\cite{Asakawa:2015ybt,Bzdak:2019pkr,Bluhm:2020mpc}.
In the present article, we shall, however, focus on the {\em dynamical}
aspects of the fluctuation of the
order parameter in the normal phase, say, above the critical temperature,
where the dynamical fluctuations  of 
the order parameter 
has a dominant strength in the {\em space-like region}.
This is the collective particle-hole excitation corresponding to 
density fluctuations, and has the nature of the soft mode of the QCD-CP
in the sense that 
the peak energy of the collective mode goes down (softens)
as the system approaches the QCD-CP in the normal phase and eventually 
becomes zero 
there~\cite{Fujii:2003bz,Fujii:2004jt,Yokota:2016tip,Yokota:2017uzu}.

As for the CSC,
which is caused by attractive diquark correlations (Cooper pairs) 
in the presence of Fermi sphere,
there are various patterns of CSC due to
the intrinsic degrees of freedom of quarks~\cite{Alford:2007xm}. 
Among such varieties of the pairing patterns,
 we only take the two-flavor color superconductivity (2SC)
in the present article,
and consider the precursory diquark fluctuations, assuming
the second-order nature of the phase transition
\footnote{It should be remarked that gluon 
fluctuations may alter the order of the 
phase transition to a (weak) first order from a second one in a relatively 
low density
~\cite{Matsuura:2003md,Giannakis:2004xt,Noronha:2006cz,Fejos:2019oxz}.
It seems, however, there are no conclusive claims on 
the strength of the first-order nature in the low-density region.
Thus, we will assume that the phase transition is second-order 
or weakly first-order in this article.}.
In fact, it was shown that there exists a specific soft mode in the normal phase 
of the CSC, 
which has a prominent spectral support in the space-like 
region~\cite{Kitazawa:2001ft,Kitazawa:2005vr,Nishimura:2022mku,Nishimura:2023oqn}. 

We shall closely examine the spectral properties of these soft modes 
associated with the respective phase transitions in a
coherent way on the basis 
of the massive 2-flavor and 3-color Nambu-Jona-Lasinio (NJL) 
model~\cite{Nambu:1961tp,Nambu:1961fr,Vogl:1991qt,Klevansky:1992qe,Hatsuda:1994pi,Ebert:1994mf,Buballa:2003qv}, basically following 
Refs.~\cite{Kitazawa:2001ft,Kitazawa:2005vr,Nishimura:2022mku,Nishimura:2023oqn,Nishimura:2024kvz}.

Then, we first show that 
the diquark soft mode of the 2SC causes the 
{\em pseudogap in the quark spectra}, i.e., a depression in 
the density of states of the quark spectra around the Fermi energy,
 above but in the vicinity of 
the critical temperature.

Although it is left as a future problem to find good observables
to confirm the pseudogap experimentally,
we shall show that the soft modes for both phase transitions 
cause an anomalous enhancement 
of electric conductivity and 
the dilepton production rate (DPR) 
to be observed by HIC.
For that, we employ an idea 
established in condensed matter physics  
to account for an anomalous excess of the electric conductivity, 
known as  
`para-conductivity'~\cite{AL:1968,Maki:1968,Thompson:1968,Larkin:book}, 
in  metal superconductors~\cite{Larkin:book,tinkham2004introduction}.
Needless to say,
the DPR should be useful to detect the 
dynamical properties of the created matter by HIC owing to
 the relatively weak interactions with the surrounding matter,
and the significance of electric conductivity in hot and dense quark matter 
~\cite{Arnold:2000dr,Arnold:2003zc,Teaney:2006nc,Cassing:2013iz,Greif:2014oia,Aarts:2020dda,Kaczmarek:2022ffn} 
has been being revealed for understanding the 
space-time evolution of the created matter 
in HIC~\cite{ Hirono:2012rt,Nakamura:2022wqr,Mayer:2024kkv}.

The present paper is organized as follows. 
In the next section, we introduce the model Lagrangian, and show the phase diagram given by it in the
mean-field approximation. 
In \S~\ref{sec:Soft-modes}, a unified account is given of the soft modes as collective excitations 
based on the linear-response theory, and
the important difference in the analytic properties of the spectral functions 
are elucidated; then, paying attention to the respective analytic properties, we
derive approximate low-energy effective propagators valid in the vicinity of the 
respective critical points.
In \S~\ref{sec:Pseudogap}, we calculate the density of states of
the quark spectra and establish the emergence of the pseudogap phenomenon.
In \S~\ref{sec:sigma,DPR}, we calculate the photon self-energy in the medium 
by taking account of the soft modes, and  
demonstrate  that the soft modes cause an anomalous enhancement
of the electric conductivity and the dilepton production rate
near but above the respective critical temperatures of the 2SC and QCD-CP.
The last section is devoted to a brief summary and concluding remark.

\section{Model Lagrangian and phase diagram}
\label{sec:Model}

To explore the effects of critical fluctuations in dense quark matter, 
we employ a simple 2-flavor NJL model 
with a nonzero current quark mass $m$, as was done in Ref.~\cite{Nishimura:2024kvz},
\begin{align}
\mathcal{L} = \bar{\psi} i (\slashed{\partial} - m) \psi 
+G_S [(\bar\psi \psi)^2 + (\bar\psi i \gamma_5 \vec{\tau} \psi)^2]
+G_D (\bar\psi i \gamma_5 \tau_2 \lambda_A \psi^C)
                       (\bar\psi^C i \gamma_5 \tau_2 \lambda_A \psi),
\label{eq:Lagrangian}
\end{align}
where $\psi(x)$ is the quark field and 
$\psi^C (x) = i \gamma_2 \gamma_0 \bar\psi^T (x)$ denotes its charge conjugation.
$\vec{\tau} = (\tau_1, \tau_2, \tau_3)$ are the Pauli matrices for the flavor $SU(2)_f$,
and $\lambda_A~(A = 2,5,7)$ are the antisymmetric components 
of the Gell-Mann matrices for the color $SU(3)_c$.
The scalar coupling constant $G_S$ and 
the three-momentum cutoff $\Lambda$ are determined 
so as to reproduce the pion mass $m_{\pi} = 138~\rm{MeV}$ and
the pion decay constant $f_{\pi} = 93~\rm{MeV}$
at the current quark mass $m = 5.5~{\rm MeV}$~\cite{Hatsuda:1994pi}:
$G_S = 5.50~\rm{GeV^{-2}}$ and  $\Lambda = 631~{\rm MeV}$.
We treat $G_D$ as a free parameter and vary it in the range obtained by various estimates~\cite{Buballa:2003qv}.

To describe the chiral restoration and the onset of the 2SC phase,
we adopt the mean-field approximation (MFA) 
for the scalar and diquark operators,
\begin{align}
\hat\sigma(\bm{x},t) = \bar\psi(\bm{x},t)\psi(\bm{x},t)
\qquad \mbox{and} \qquad 
\hat{\delta}_A(\bm{x},t) = \bar\psi^C(\bm{x},t) i \gamma_5 \tau_2 \lambda_A \psi(\bm{x},t) .
\label{eq:delta-sigma}
\end{align}
We refer to their expectation values $\langle \hat\sigma \rangle$ 
and $\langle \hat\delta_A \rangle$ as the chiral and diquark condensates, respectively. 
The Lagrangian density in this approximation takes the form
\begin{align}
\mathcal{L}_{\rm MFA} = \bar{\psi} i (\slashed{\partial} - m) \psi 
- M \bar\psi\psi - \frac12 ( \Delta^\dagger \bar\psi^C i \gamma_5 \tau_2 \lambda_A \psi + {\rm h.c.})
-\frac{M^2}{4G_S} - \frac{|\Delta|^2}{4G_D} ,
\label{eq:L_MFA}
\end{align}
with $M=-2G_S\langle \hat\sigma \rangle$ and $\Delta = -2G_D \langle \hat\delta_A \rangle$. 

From Eq.~\eqref{eq:L_MFA}, the thermodynamic potential per unit volume at temperature $T$ and 
quark chemical potential $\mu$ is calculated to be~\cite{Kitazawa:2005vr}
\begin{align}
\omega_{\rm MFA} 
&= 
\frac{(M-m)^2}{4G_S} + \frac{|\Delta|^2}{4G_D}
-4 \int \frac{d^3p}{(2\pi)^3} 
\bigg\{ E_{\bm{p}} + T{\rm log} 
\big( 1 + e^{-\xi_+ / T} \big) \big( 1 + e^{-\xi_- / T} \big) \nonumber \\
& \mathrel{\phantom{=}} + \epsilon_+ + {\rm sgn}(\xi_-)\epsilon_- + 2T {\rm log} \big( 1 + e^{-\epsilon_+ / T} \big) 
\big( 1 + e^{- {\rm sgn}(\xi_-) \epsilon_- / T} \big)
\bigg\},
\label{eq:omega_MFA} \\
&\hspace{0.3cm}
E_{\bm{p}} = \sqrt{\bm{p}^2 + M^2}, 
\qquad
\xi_\pm = E_{\bm{p}} \pm \mu ,
\qquad
\epsilon_\pm = \sqrt{\xi_\pm^2 + |\Delta|^2} .
\end{align}
The expectation values $M$ and $\Delta$ are given by minimizing $\omega_{\rm MFA}$, and 
the stationary condition gives the gap equations
\begin{align}
   \frac{\partial \omega_{\rm MFA} }{\partial M }= 0 , 
   \qquad
   \frac{\partial \omega_{\rm MFA} }{\partial \Delta }= 0 .
   \label{eq:gapeq}
\end{align}

The 2SC phase is characterized by a nonzero diquark condensate $\Delta$. 
At the 2SC-PT, $\Delta$ gets to have a non-zero value in the 2SC phase continuously from zero,
provided that the phase transition is of the second order. This implies that
 $\omega_{\rm MFA}$ satisfies
\begin{align}
    \frac{\partial^2\omega_{\rm MFA}}{\partial\Delta^2}\Big|_{\Delta=0} =0 ,
    \label{eq:d2om/dD2}
\end{align}
at the 2SC-PT.
Equation~\eqref{eq:d2om/dD2} tells us that  the thermodynamic potential around the minimum point 
is flat, and hence the diquark susceptibility 
$\chi_D = (\partial^2\omega_{\rm MFA}/\partial\Delta^2)^{-1}$ is divergent, 
and accordingly so do the fluctuations of $\Delta$  at the 2SC-PT. 
Such divergences are a general feature of second-order phase transitions~\cite{Asakawa:2015ybt}. 
We will discuss their consequences in subsequent sections. 
Owing to the nonzero current quark mass, the chiral condensate $\langle\hat\sigma\rangle$ is 
always nonzero. When $\omega_{\rm MFA}$ has two local minima, 
the values of $M$ and $\Delta$ at the global minimum can exhibit a discontinuous change 
when  $T$ and $\mu$ are varied, which 
 corresponds to a first-order phase transition. 

\begin{figure}[t]
\centering
\includegraphics[width=0.5\textwidth]{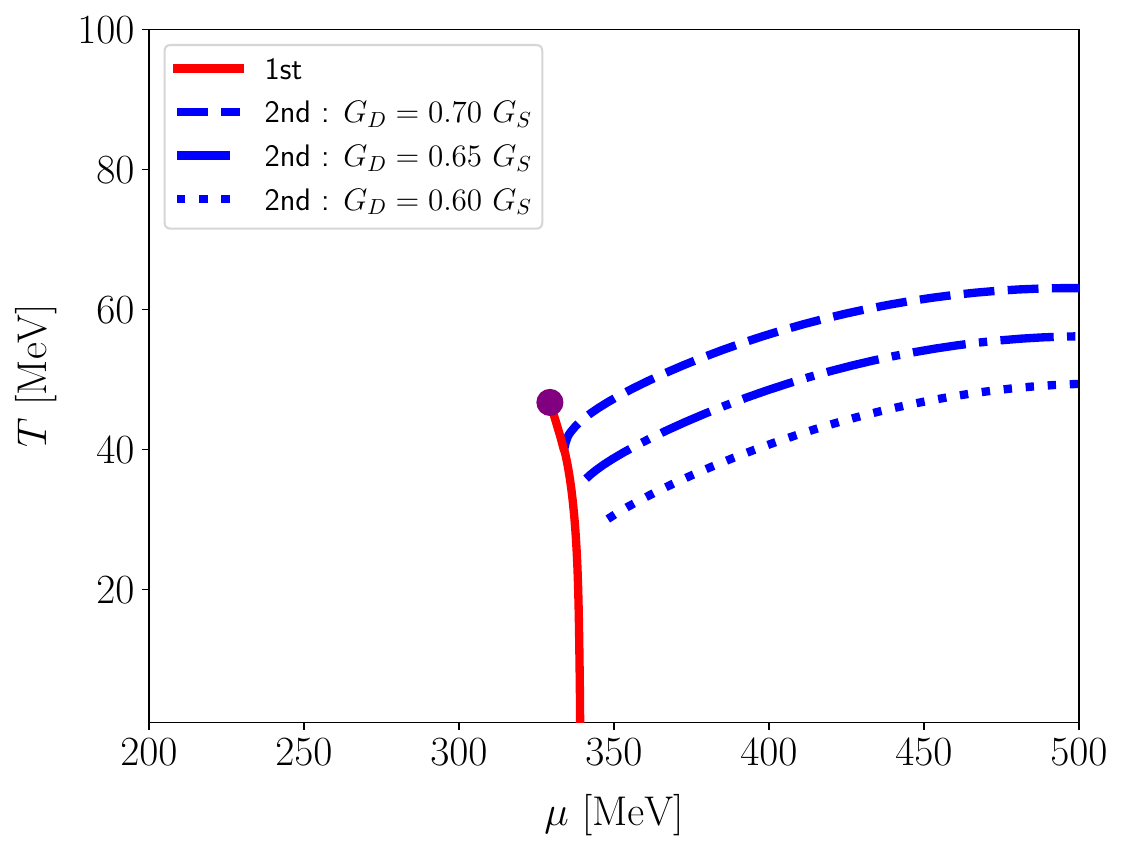}
\caption{
Phase diagram calculated by the mean-field approximation 
in the 2-flavor NJL model (\ref{eq:Lagrangian})~\cite{Nishimura:2024kvz}. 
The solid line shows the first-order phase transition calculated with $G_D=0.70G_S$.
The dashed, dash-dotted, and dotted lines are the second-order 2SC-PT 
for $G_D/G_S = 0.70$, $0.65$, and $0.60$, respectively.
The QCD-CP is indicated by the circle marker 
located at $(T_{\rm CP}, \mu_{\rm CP}) \simeq (46.712, 329.34)$~MeV.
}
\label{fig:Phase-D}
\end{figure}

In Fig.~\ref{fig:Phase-D}, we show the phase diagram in the $T$--$\mu$ plane 
obtained by the MFA~\cite{Nishimura:2024kvz}.
The solid line shows the first-order transition line.
The circle marker at $(T_{\rm CP},\, \mu_{\rm CP}) \simeq (46.712,\, 329.34)~{\rm MeV}$ 
denotes the QCD-CP, which is an endpoint of the first-order transition line where the phase transition is of second order.
The dashed, dash-dotted, and dotted lines show the 2SC-PT for $G_D/G_S = 0.70$, $0.65$, and $0.60$, respectively.
The critical temperature of the 2SC-PT increases as $G_D$ becomes larger.
The 2SC-PT is of second order in our model.

As mentioned in Introduction, several studies beyond the MFA suggest that the order of the 2SC-PT is a weak first order owing to fluctuations of gluon fields~\cite{Matsuura:2003md,Giannakis:2004xt,Fejos:2019oxz}.
Since any global symmetry distinguishes
the 2SC phase from the normal one, 
the transition to the 2SC phase may have to be crossover~\cite{Alford:2007xm}.
In any event, we are not aware of a definite conclusion on the order 
of the 2SC-PT,
the following analyses will be performed simply based on the result of the MFA.

At the QCD-CP where two minima of $\omega_{\rm MFA}$ existing in the lower-$T$ and higher-$\mu$ region merge, 
$\omega_{\rm MFA}$ has a flat direction on the $M$--$\Delta$ plane. 
As Fig.~\ref{fig:Phase-D} shows, in our model $\Delta=0$ is satisfied at the QCD-CP for $G_D/G_S<0.7$ 
and the flat direction is along the $M$ direction. Therefore, at the minimum $\omega_{\rm MFA}$ satisfies
\begin{align}
    \frac{\partial^2\omega_{\rm MFA}}{\partial M^2} = 0 ,
    \label{eq:d2om/dM2}
\end{align}
at the QCD-CP. Equation~\eqref{eq:d2om/dD2} shows that the fluctuation of $M$ 
is divergent at the QCD-CP.

\section{Collective diquark/particle-hole excitations as the soft modes of the phase transitions} 
\label{sec:Soft-modes}

In this section, 
we discuss the dynamical properties of fluctuations of $\Delta$ and $M$ 
near the 2SC-PT and QCD-CP, respectively, based on the linear response theory.
We show that these fields exhibit collective excitations with prominent peaks of the 
strength function near these phase transitions, 
respectively~\cite{Kitazawa:2001ft,Kitazawa:2005vr,Fujii:2003bz,Fujii:2004jt,Nishimura:2024kvz}.

\subsection{Linear response theory}
\label{sec:LRT}

The linear-response theory~\cite{fetter2012quantum} is a useful tool to explore dynamical 
properties of collective excitations. A key idea of this theory is to disturb the system with an infinitesimal external field represented by the Hamiltonian $H_{\rm ext}=\int d^3xdt e^{i\omega t-i\bm{k}\cdot\bm{x}} f(\bm{x},t) {\cal O}(\bm{x},t)$, where ${\cal O}(\bm{x},t)$ is a bosonic-field operator and $f(\bm{x},t)$ is a classical function. As a result of applying the external field, the expectation value of ${\cal O}(\bm{x},t)$ can deviate from its thermal expectation value $\langle{\cal O}\rangle$. For an infinitesimal perturbation, this deviation is proportional to $f(\bm{x},t)$ and represented by~\cite{fetter2012quantum}
\begin{align}
    \delta \langle{\cal O}(\bm{x},t)\rangle
    \equiv 
    \langle{\cal O}(\bm{x},t)\rangle_{\rm ext} - \langle{\cal O}\rangle = \int d^3x'dt' D^R(\bm{x}-\bm{x}',t-t') f(\bm{x}',t'),
    \label{eq:LRT1}
\end{align}
where $\langle{\cal O}(\bm{x},t)\rangle_{\rm ext}$ represents the expectation value with the external field and $D^R(\bm{x},t)$ is the retarded Green's function 
\begin{align}
    D^R(\bm{x},t) =&\ -i \langle [ {\cal O}(\bm{x},t),{\cal O}(\bm{0},0)] \rangle \theta(t),
    \label{eq:Xi^R(x)}
\end{align}
with $[A,\,B]=AB-BA$ being the commutator.
The Fourier transformation of Eq.~\eqref{eq:LRT1} leads to
\begin{align}
    \delta \langle{\cal O}(\bm{k},\omega)\rangle = D^R(\bm{k},\omega) f(\bm{k},\omega),
    \label{eq:LRT2}
\end{align}
with $D^R(\bm{k},\omega) = \int d^3xdt e^{i\omega t - i\bm{k}\cdot\bm{x}} D^R(\bm{x},t)$ and so on.

When $D^R(\bm{k},\omega)$ has a pole at $\omega=\omega(\bm{k})$ for real $\omega$, Eq.~\eqref{eq:LRT2} tells us that $\delta \langle{\cal O}(\bm{k},\omega(\bm{k}))\rangle$ becomes nonzero with an infinitesimal external perturbation. Such a mode forms a collective excitation that 
carries the quantum number of 
the operator ${\cal O}(\bm{x},t)$. When $D^R(\bm{k},\omega)$ admits a complex pole $\omega=\omega(\bm{k})\in\mathbb{C}$ with a small imaginary part, it is also said that there exists a well-developed collective mode or quasi-particle.

To explore the fluctuations of $\Delta$ and $M$, we only have to substitute the operators $\hat\delta_A(\bm{x},t)$ and $\hat\sigma(\bm{x},t)$ into $\hat{\cal O}(\bm{x},t)$, respectively. We denote these retarded functions with ${\cal O}(\bm{x},t)=\hat\delta_A(\bm{x},t)$ and $\hat\sigma(\bm{x},t)$ as $D^R_D(\bm{k},\omega)$ and $D^R_S(\bm{k},\omega)$, respectively.

\begin{figure}[t]
\centering
\includegraphics[keepaspectratio, scale=0.5]{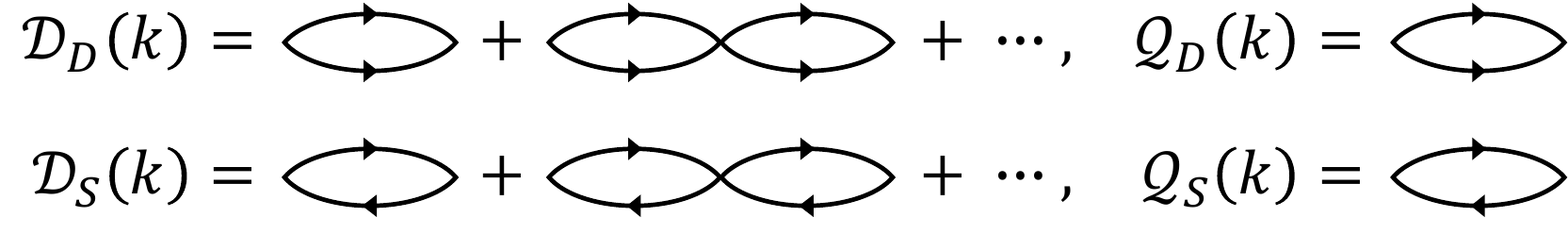}
\caption
{
Diagrammatic representation of Eq.~(\ref{eq:D}).
The lines denote the quark propagator.
}
\label{fig:D-Q}
\end{figure}

\subsection{Random-phase approximation and Thouless criterion}
\label{sec:RPA}

The retarded Green's functions $D^R_D(\bm{k},\omega)$ and $D^R_S(\bm{k},\omega)$ that are consistent with the MFA are given by the random-phase approximation (RPA) or ring approximation, 
\begin{align}
    D^R_\gamma(\bm{k},\omega) = \frac{Q^R_\gamma(\bm{k},\omega)}{ 1 + G_\gamma Q^R_\gamma(\bm{k},\omega)}
    = Q^R_\gamma(\bm{k},\omega) - Q^R_\gamma(\bm{k},\omega) G_\gamma Q^R_\gamma(\bm{k},\omega) + \cdots,
    \label{eq:D}
\end{align}
with $\gamma=D,S$ and the unperturbed correlation functions $Q^R_\gamma(\bm{k},\omega)$. As shown in Fig.~\ref{fig:D-Q}, $Q^R_\gamma(\bm{k},\omega)$ are represented by the one-loop graphs, where the directions of quark propagators are different for $\gamma=D,S$.
To calculate $Q^R_\gamma(\bm{k},\omega)$, it is convenient to first introduce the corresponding functions in the Matsubara (imaginary-time) formalism
\begin{align}
    {\cal Q}_D(k) 
    =&\ \int d^3xdt e^{i\nu_n \tau-i\bm{k}\cdot\bm{x}}
    \langle T_\tau \hat\delta_A(\bm{x},\tau) \hat\delta_A^\dagger(\bm{0},0) \rangle_{\rm free}
    \notag \\
    =&\ -2N_f(N_c-1) \int_p {\rm Tr}_D[ {\cal G}_0(p) {\cal G}_0(k-p)],
    \label{eq:Q_Delta}
    \\
    {\cal Q}_S(k) 
    =&\ \int d^3xdt e^{i\nu_n \tau-i\bm{k}\cdot\bm{x}}
    \langle T_\tau \hat\sigma(\bm{x},\tau)\hat\sigma(\bm{0},0) \rangle_{\rm free}
    \notag \\
    =&\ -2N_f N_c \int_p {\rm Tr}_D [ {\cal G}_0(p) {\cal G}_0(k+p)],
    \label{eq:Q_S}
\end{align}
where $k=(\bm{k},i\nu_n)$ is the collective index with $\nu_n=2n\pi/T$ the Matsubara frequency for bosons, 
$\langle\cdot\rangle_{\rm free}$ denotes the expectation value 
in the non-interacting system, and
$\mathcal{G}_0 (p) = \mathcal{G}_0 (\bm{p}, i\omega_m) 
= {1}/[(i\omega_m + \mu) \gamma_0 - \bm{p} \cdot \bm{\gamma}-M]$
is the free-quark propagator with $\omega_m=(2m+1)\pi/T$ 
being the Matsubara frequency for fermions.
The retarded functions $Q^R_\gamma(\bm{k},\omega)$ is then obtained by the analytic continuation $i\nu_n \rightarrow \omega+i\eta$
\begin{align}
    {\cal Q}_\gamma (k) 
    = {\cal Q}_\gamma (\bm{k},i\nu_n) \xrightarrow[i\nu_n \rightarrow \omega+i\eta]{} Q^R_\gamma(\bm{k},\omega) .
\end{align}

For later convenience, we also introduce the retarded $T$-matrices
\begin{align}
\Xi^R_\gamma(\bm{k},\omega) 
=& \frac1{G_\gamma^{-1}+Q^R_\gamma(\bm{k},\omega)}
=  G_\gamma - G_\gamma D^R_\gamma(\bm{k},\omega) G_\gamma,
\label{eq:genf-T-matrix}
\end{align}
which also means
\begin{align}
  D^R_\gamma(\bm{k},\omega)=G_\gamma^{-1} Q^R_\gamma (\bm{k},\omega)\Xi^R_\gamma (\bm{k},\omega).  
\end{align}

An important property of the $T$-matrices~\eqref{eq:genf-T-matrix} is that their low-energy low-momentum limits are related to the second derivatives of $\omega_{\rm MFA}$ as~\cite {Nishimura:2024kvz}
\begin{align}
\lim_{|\bm{k}|\to0} {\Xi^R_D}^{-1}(\bm{k}, 0) =
2 \frac{\partial^2 \omega_{\rm MFA}}{\partial \Delta^2},
\qquad
\lim_{|\bm{k}|\to0} {\Xi^R_S}^{-1}(\bm{k}, 0) = 2 \frac{\partial^2 \omega_{\rm MFA}}{\partial M^2} .
\label{eq:Thouless}
\end{align}
Since the thermodynamic potential satisfies
$\partial^2 \omega_{\rm MFA}/\partial\Delta^2=0$ ($\partial^2 \omega_{\rm MFA}/\partial M^2=0$) at the 2SC-PT (QCD-CP) as discussed in Sec.~\ref{sec:Model}, from Eq.~\eqref{eq:Thouless} it is immediately concluded that 
\begin{align}
{\Xi^R_\gamma}^{-1}(\bm{0}, 0) =0 
\quad \mbox{and} \quad
{D^R_\gamma}^{-1}(\bm{0}, 0) =0 ,
\label{eq:Thouless_gen}
\end{align}
are satisfied at the respective critical points.
These properties are called the Thouless criterion~\cite{Thouless}.
Although the derivation of the Thouless criterion presented here relies on the MFA and the RPA, it has 
a general validity beyond the MFA, reflecting the fact that $\lim_{|\bm{k}|\to0} {\Xi^R_\gamma}^{-1}(\bm{k}, 0)$ corresponds to the susceptibility of the order-parameter field that diverges at the second-order 
phase transition~\cite{Larkin:book}. 

An important consequence of the Thouless criterion is that $D^R_D(\bm{k}, \omega)$ and $D^R_S(\bm{k}, \omega)$ have a 
{\em massless pole} $\omega(\bm{0})=0$ at the 2SC-PT and QCD-CP, respectively. Since the location of the pole $\omega(\bm{k})$ changes continuously as a function of $T$ and $\mu$, the pole stays near the origin $\omega=0$ even away from the transition point. Such a mode is called the {\em soft mode} of the respective phase transitions.

\subsection{Analytic structure of $Q^R_\gamma$}
\label{sec:Q}

The imaginary parts of $Q^R_\gamma(\bm{k},\omega)$ are calculated to be~\cite{Nishimura:2024kvz}
\begin{align}
{\rm Im} Q^R_D (\bm{k}, \omega) 
=& -\frac{N_f (N_c - 1) T}{4 \pi} \frac{(\omega + 2\mu)^2 - \bm{k}^2}{|\bm{k}|}
\nonumber \\
&\times
\bigg\{
\theta \bigl( \bar\Lambda - |\omega+2\mu| \bigr) 
\theta \bigl( |\omega+2\mu| - \sqrt{\bm{k}^2 + 4M^2} \bigr) 
F_D \bigl( \omega, \bar{k}(|\bm{k}|, \omega+2\mu) \bigr)
\nonumber \\ 
&\qquad +
\theta \bigl( \bar{k}(|\bm{k}|, \bar\Lambda) - |\omega+2\mu| \bigr) 
\Big[ 
F_D \bigl( \omega, \bar{k}(|\bm{k}|, \omega+2\mu) \bigr) - 
F_D \bigl( \omega, \bar\Lambda \bigr)
\Big]
\bigg\},
\label{eq:ImQ_2SC}
\\
F_D (\omega, x) =& \ 2 \sum_{s=\pm} s~{\rm log}~{\rm cosh} ([\omega+sx]/4T) ,
\end{align}
and
\begin{align}
{\rm Im} Q^R_S (\bm{k}, \omega) 
=& -\frac{N_f N_c T}{4 \pi} \frac{\omega^2 - \bm{k}^2 - 4M^2}{|\bm{k}|}
\nonumber \\
&\times
\bigg\{
\theta \bigl( \bar\Lambda - |\omega| \bigr) 
\theta \bigl( |\omega| - \sqrt{\bm{k}^2+4M^2} \bigr) 
F_S \bigl( \omega, \bar{k}(|\bm{k}|, \omega) \bigr)
\nonumber \\ 
&\qquad +
\theta \bigl( \bar{k}(|\bm{k}|, \bar{\Lambda}) - |\omega| \bigr) 
\Big[ 
F_S \bigl( \omega, \bar{k}(|\bm{k}|, \omega) \bigr) - 
F_S \bigl( \omega, \bar{\Lambda} \bigr)
\Big]
\bigg\},
\label{eq:ImQ_QCDCP} 
\\
F_S (\omega, x) =& \sum_{s, t=\pm} s~{\rm log}~{\rm cosh} ([\omega + sx - 2t\mu]/4T) ,
\end{align}
with $\bar{k}(|\bm{k}|, \omega) =\ |\bm{k}| \sqrt{1 - 4M^2 / (\omega^2-\bm{k}^2)}$ and $\bar\Lambda = 2\sqrt{\Lambda^2+M^2}$.

From Eqs.~\eqref{eq:ImQ_2SC} and~\eqref{eq:ImQ_QCDCP}, one finds that the first (second) term in the curly bracket in Eq.~(\ref{eq:ImQ_2SC}) takes a nonzero value at 
\begin{align}
  |\omega + 2\mu| > \sqrt{\bm{k}^2 + 4M^2}, \quad \quad 
  (|\omega + 2\mu| < \bar{k}(|\bm{k}|, \bar\Lambda)),  
  \label{eq:supprtD}
\end{align}
while that in Eq.~(\ref{eq:ImQ_QCDCP}) is nonzero at 
\begin{align}
  |\omega| > \sqrt{\bm{k}^2 + 4M^2}, \quad \quad
(|\omega| < \bar{k}(|\bm{k}|, \bar\Lambda)).
\label{eq:supprtS}
\end{align}
One can also verify from Eqs.~\eqref{eq:ImQ_2SC} and~\eqref{eq:ImQ_QCDCP} that $Q_D^R(\bm{k},\omega)$ and $Q_S^R(\bm{k},\omega)$ are not analytic at the boundary of the supports~\eqref{eq:supprtD} and~\eqref{eq:supprtS}.
Therefore, $Q_S^R(\bm{k},\omega)$ is not analytic at the origin, whereas $Q_D^R(\bm{k},\omega)$ is analytic there. As we will see later, this leads to a qualitative difference in the nature of the soft modes of the 2SC-PT and QCD-CP.

\subsection{Linearized time-dependent Ginzburg-Landau (TDGL) approximation}
\label{sec:TDGL}

In this subsection, we derive effective equations for the soft modes near the 2SC-PT and QCD-CP known as the linearized time-dependent Ginzburg-Landau (TDGL) equations~\cite{Nishimura:2024kvz}. These equations are obtained by expanding $\Xi_\gamma^R(\bm{k},\omega)$ with respect to $\bm{k}$ and $\omega$. 
Through the derivation of these equations, the qualitative difference in the analytic properties of the soft modes of the 2SC-PT and QCD-CP will be clarified.
The resultant TDGL equations will be found helpful to 
investigate the effects of the soft modes on various observables
in a simple way to be done in the subsequent sections.

\subsubsection{Soft mode of 2SC-PT}
\label{sec:TDGL:2SC}

Let us start with the soft mode of the 2SC-PT, which is a collective mode encoded in ${\Xi^R_D} (\bm{k}, \omega)$. As discussed in Sec.~\ref{sec:RPA}, this $T$-matrix satisfies ${\Xi^R_D}^{-1} (\bm{0},0)=0$ at $T=T_c$ and is analytic at $\omega=|\bm{k}|=0$. Therefore, at small $\omega$ and $|\bm{k}|$ this function is well approximated by the Taylor expansion 
\begin{align}
{\Xi^R_D}^{-1} (\bm{k}, \omega) 
\simeq A_D (\bm{k}) + C_D \omega,
\label{eq:Xi-LEE_2SC}
\end{align}
near the 2SC-PT with
\begin{align}
    A_D (\bm{k}) = G_D^{-1} + Q^R_D (\bm{k}, 0) , \qquad 
    C_D = \frac{\partial Q^R_D (\bm{0}, \omega)}{ \partial \omega} \bigg|_{\omega=0},
\label{eq:Def-A_D-C_D}
\end{align}
which are found to be real and complex numbers, respectively. Since the Thouless criterion~\eqref{eq:Thouless} tells us that 
$A_D(\bm{0})=0$ at $T=T_c$, Eq.~\eqref{eq:Xi-LEE_2SC} is further expanded as
\begin{align}
{\Xi^R_D}^{-1} (\bm{k}, \omega)
\simeq \tilde a_D \epsilon + b_D \bm{k}^2 + c_D \omega ,
\label{eq:Xi-TDGL_2SC} 
\end{align}
with the reduced temperature 
\begin{align}
    \epsilon = \frac{T-T_c}{T_c} .
    \label{eq:reduced-T}
\end{align}
The approximate formula~\eqref{eq:Xi-TDGL_2SC} corresponds to the linearized time-dependent Ginzburg-Landau (TDGL) equation~\cite{Larkin:book}.
In fact, the linear-response theory, Eq.~\eqref{eq:LRT2}, shows 
that the equation of motion of the field $\Delta$ with an infinitesimal external field is 
given by ${\Xi^R_D}^{-1} (\bm{k}, \omega)\Delta(\bm{k}, \omega)=0$, 
whose Fourier transformation gives the linearized TDGL equation 
$(iC_D \partial/\partial\omega - b_D \nabla^2 + \tilde a_D\epsilon) \Delta=0$.
In the following, we refer to Eqs.~\eqref{eq:Xi-TDGL_2SC} and~(\ref{eq:Xi-LEE_2SC}) as the TDGL and the low-energy (LE) approximations, respectively.
It is numerically verified that these approximations well reproduce the ${\Xi^R_D}^{-1} (\bm{k}, \omega)$ obtained in the RPA near the 2SC-PT~\cite{Nishimura:2024kvz}.

From Eq.~\eqref{eq:Xi-TDGL_2SC}, the dispersion relation of the soft mode is readily obtained as
\begin{align}
    \omega=-(\tilde a_D\epsilon+b_D\bm{k}^2)/c_D .
\end{align}
When $|{\rm Re}c_D|\ll|{\rm Im}c_D|$, Eq.~\eqref{eq:Xi-TDGL_2SC} 
can be rewritten as
\begin{align}
\Xi^R_D (\bm{k}, \omega) 
= \frac{1}{a_D + b_D \bm{k}^2 + c_D \omega}
=\frac{i}{|{\rm Im} c_D|}
\frac{1}{\omega +i \tau_{\rm GL}^{-1}
\big(
1 + \xi_D^2 \bm{q}^2
\big)},
\label{eq:TDGLtauxi}
\end{align}
where $\tau_{\rm GL}=|c_D|/a_D$ and $\xi_D=\sqrt{b_D/a_D}$ are the relaxation 
time and the coherence length of the soft mode, respectively.
Equation~\eqref{eq:TDGLtauxi} shows that the soft mode is a damping mode.

Using the approximations $M \ll T+|\mu| \ll \Lambda$ and $T/\mu\ll1$, the coefficients in Eq.~\eqref{eq:Xi-TDGL_2SC} are calculated to be
\begin{align}
\tilde{a}_D = \frac{2N_f(N_c-1)}{\pi^2} \mu^2,
\quad
b_D = \frac{7N_f(N_c-1)\zeta(3)}{48\pi^4} \frac{\mu^2}{T^2},
\quad
c_D = -i\frac{N_f(N_c-1)}{4\pi} \frac{\mu^2}T,
\label{eq:abc_simplified-TDGL}
\end{align}
up to logarithmic terms~\cite{Nishimura:2024kvz}.
On account of Eq.~\eqref{eq:abc_simplified-TDGL}, the expressions of the 
relaxation time and the coherence length in Eq.~\eqref{eq:TDGLtauxi} 
are simplified to
\begin{align}
\tau_{\rm GL} = \frac\pi{8T} \frac1\epsilon,
\qquad
\xi_D = \sqrt{\frac{7\xi(3)}{96T^2}} \, \frac1{\epsilon^{1/2}},
    \label{eq:tau-xi_2SC}
\end{align}
respectively.
One sees that 
the both quantities do not depend on $\mu$ and are divergent for $\epsilon \rightarrow 0$.

\subsubsection{Soft mode of QCD-CP}
\label{sec:TDGL:QCDCP}

The soft mode of the QCD-CP can be treated in much the same way as the 2SC-PT. 
$\Xi^R_S (\bm{k}, \omega)$ is approximated for small $\omega$ and $|\bm{k}|$ as
\begin{align}
{\Xi^R_S}^{-1} (\bm{k}, \omega) 
\simeq A_S(\bm{k}) + C_S(\bm{k}) \omega,
\label{eq:Xi-LEE_QCDCP}
\end{align}
with 
\begin{align}
A_S(\bm{k}) = G_S^{-1} + Q^R_S(\bm{k}, 0) \quad
\textrm{and} \quad
C_S(\bm{k}) = 
\frac{\partial Q^R_S (\bm{k}, \omega) }{ \partial \omega }\bigg|_{\omega=0},
\label{eq:Def-A_S-C_S}
\end{align}
which are found to be real and pure imaginary, respectively.
An important difference of Eq.~\eqref{eq:Def-A_S-C_S} from Eq.~\eqref{eq:Xi-LEE_2SC} is 
that ${\Xi^R_S}^{-1} (\bm{k}, \omega)$ is not analytic at $\omega=|\bm{k}|=0$; 
reflecting it, $C_S(\bm{k})$ diverges as $1/|\bm{k}|$ for $\bm{k} \rightarrow \bm{0}$.
As discussed in Sec.~\ref{sec:Q}, $\Xi_S^R(\bm{k},\omega)$ has discontinuities at $|\omega|=\bar k(|\bm{k}|,\bar\Lambda)\simeq|\bm{k}|$. Therefore, the LE approximation~\eqref{eq:Xi-LEE_QCDCP} is 
valid only in the region $|\omega|<\bar k(|\bm{k}|,\bar\Lambda)$ and not applicable to the time-like region.
This analytic property shows that the soft mode of the QCD-CP is in the {\em space-like region},
 and hence not conventional mesonic excitations in the time-like region.

Equation~\eqref{eq:Xi-LEE_QCDCP} is further expanded as 
\begin{align}
{\Xi^R_S}^{-1} (\bm{k}, \omega) 
\simeq \ a_S(T,\mu) + b_S|\bm{k}|^2 + c_S \frac\omega{|\bm{k}|},
\label{eq:Xi-TDGL_QCDCP}
\end{align}
with $a_S(T,\mu)=G_S^{-1} + \lim_{|\bm{k}|\to0} Q^R_S(\bm{k}, 0)$.
We refer to Eq.~\eqref{eq:Xi-TDGL_QCDCP} as the TDGL approximation for $\Xi^R_S(\bm{k},\omega)$ in analogy with Eq.~\eqref{eq:Xi-TDGL_2SC}. 
The parameter $a_S(T,\mu)$ in Eq.~\eqref{eq:Xi-TDGL_QCDCP} vanishes at the QCD-CP as in the case of 2SC-PT. 
Its behavior around there is, however, more intricate~\cite{Nishimura:2024kvz}:
When $T$ and $\mu$ approach the QCD-CP linearly but with a different fixed ratio
 $T-T_{\rm CP}:\mu-\mu_{\rm CP}$, it varies as 
\begin{align}
a_S \sim
\begin{cases}
~\epsilon_{\rm CP} & \text{parallel to the first-order line,} \\
~\epsilon_{\rm CP}^{2/3} & \text{otherwise,} 
\end{cases}
\label{eq:a_S}
\end{align}
in the MFA with 
\begin{align}
\epsilon_{\rm CP} = \sqrt{
~ \bigg( \frac{T - T_{\rm CP}}{T_{\rm CP}} \bigg)^2 
+ \bigg( \frac{\mu - \mu_{\rm CP}}{\mu_{\rm CP}} \bigg)^2~
}.
\end{align}

In contrast to the case for the 2SC-PT, Eqs.~\eqref{eq:Xi-LEE_QCDCP} or~\eqref{eq:Xi-TDGL_QCDCP} is applicable only to the space-like region. When we use it, we thus assume ${\Xi^R_S} (\bm{k}, \omega) =0$ in the time-like region. The last term in Eq.~\eqref{eq:Xi-TDGL_QCDCP} is finite in the space-like region.
It is also worth mentioning that the soft mode of the QCD-CP existing in the space-like region should not be confused with the mesonic excitation in the time-like region, which is usually called the $\sigma$ meson. 
Therefore,  one should be cautious, 
when the symbol $\sigma$ is used to represent the soft mode of the QCD-CP in the literature.

\section{Emergence of pseudogap in quark excitation spectra}
\label{sec:Pseudogap}

In the previous sections, we have seen that the fluctuations of $\Delta$ and $M$ are 
enhanced near the 2SC-PT and QCD-CP, respectively, 
and  well-developed collective modes, the soft modes, are formed, which become
 massless at the critical points.
The emergence of the soft modes can, in turn, modify properties of 
various physical observables near the transition points. 
In this and the next sections, 
we investigate some such observables in the dense quark matter near the 2SC-PT and QCD-CP.

In this section, we focus on the modification of the excitation properties of quarks 
due to the soft modes.
In the strongly correlated superconductors, 
such as the high-temperature superconductors and cold atoms near the unitarity limit,
it is known that there appear unconventional properties in the fermionic excitations 
near but above the critical temperature $T_c$. They lead to the suppression of the density of states (DOS)
near the Fermi surface even above the critical temperature, which is called the pseudogap phenomenon. 
In this section, we explore the possibility of the appearance of the pseudogap in the quark spectral 
function near the 2SC-PT and QCD-CP~\cite{Kitazawa:2003cs,Kitazawa:2013zya,Kitazawa:2014sga}. 

The excitation properties of quarks are represented by the one-particle quark spectral function,
\begin{align}
{\cal A}( \bm{k},\omega )
=-\frac1\pi \cdot{\rm Im}G^R ( \bm{k},\omega )
= -\frac1\pi \frac{ G^R( \bm{k},\omega ) 
- \gamma^0 G^{R\dag}( \bm{k},\omega ) \gamma^0 }{2i},
\end{align}
with the retarded quark Green's function $G^R( \bm{k},\omega )$.
From rotational and parity invariance, the Dirac indices of ${\cal A}( \bm{k},\omega )$ can be decomposed into 
\begin{align}
{\cal A}( \bm{k},\omega )=
\rho_0( \bm{k},\omega ) \gamma^0 
- \rho_{\rm v}( \bm{k},\omega ) \hat{\bm{k}}\cdot\bm\gamma +
\rho_{\rm s}( \bm{k},\omega ),
\end{align}
with $\hat{\bm{k}} = \bm{k}/|\bm{k}|$.
Here, $\rho_0( \bm{k},\omega )$ represents the strength of excitations carrying the quark number, and the DOS of the quarks is defined through this channel as
\begin{align}
N(\omega) = 4\int \frac{d^3 {\bm{k}}}{(2\pi)^3}
{\rm Tr}_{\rm c,f}\left[ \rho_{0}({\bm{k}},\omega) \right],
\label{eqn:N_0}
\end{align}
with ${\rm Tr}_{\rm c,f}$ 
denoting the trace over color and flavor indices.

\begin{figure}[tb]
\begin{center}
\includegraphics[width=12cm]{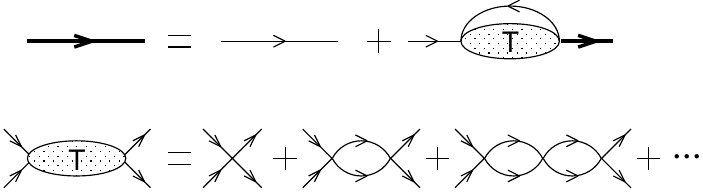} 
\caption{
Feynman diagrams representing the quark Green function in the non-self-consistent $T$-matrix approximation.
The thin lines represent the free propagator ${\cal G}_0$,
while the bold ones represent the full propagator ${\cal G}$.
}
\label{fig:quark-selfE}
\end{center} 
\end{figure}

To calculate the quark Green's function by incorporating the effects of the soft modes of the 2SC-PT, we employ the non-self-consistent $T$-matrix approximation~\cite{Kitazawa:2003cs,Kitazawa:2005vr}, where the quark propagator is diagrammatically represented as in Fig.~\ref{fig:quark-selfE}.
The thin and bold lines in the figure represent the free and full propagators.
In the Matsubara formalism, the quark self-energy in this approximation is given by
\begin{align}
\tilde\Sigma( \bm{p},\omega_n ) 
= -4 \sum_{ A=2,5,7 } ( \lambda_A )^2
T\sum_m \int \frac{d^3 \bm{k} }{(2\pi)^3}
\tilde\Xi_D ( \bm{p}+\bm{k}, \omega_n+\omega'_m )
{\cal G}_0 ( \bm{k},\omega'_m ) ,
\label{eqn:SE1}
\end{align}
where $\tilde\Xi_D ( \bm{p}, \nu_n )$ is the $T$-matrix in the imaginary-time formalism.
In Eq.~\eqref{eqn:SE1}, effects of the diquark soft modes are
contained in the propagation of quarks through the $T$-matrix.

\begin{figure}[tbp]
\begin{center}
\includegraphics[width=14cm]{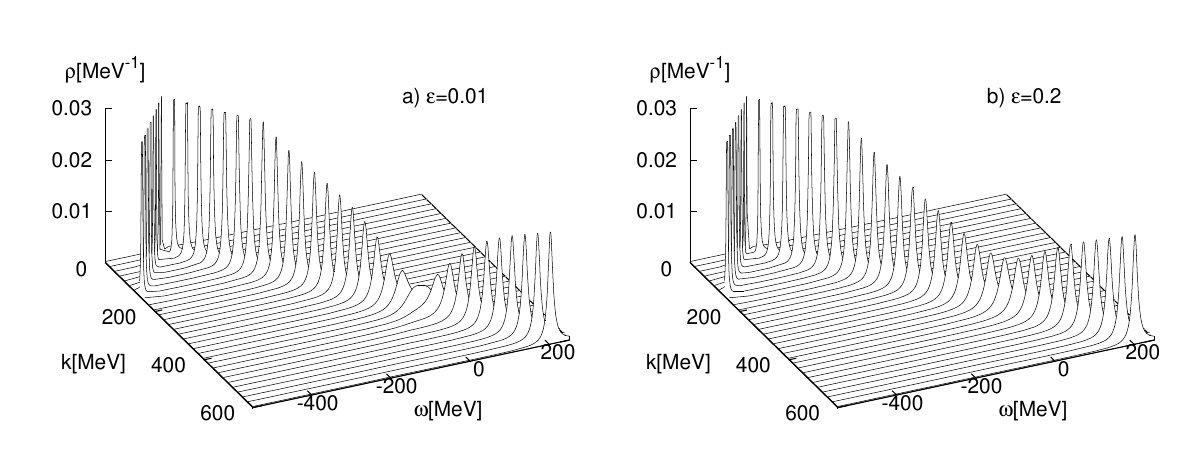}
\caption{
Spectral function $\rho_0(\bm k,\omega)$ at $\mu=400$~MeV
and $ \varepsilon=0.01 $ and $0.2$.
The peaks at $ \omega=k-\mu $ and $ \omega=-k-\mu$ correspond
to the quark and anti-quark quasiparticles, respectively.
Notice that
there is a depression around $ \omega=0 $,
which is responsible for the pseudogap formation.
}
\label{fig:spc}
\end{center} 
\end{figure}

In Fig.~\ref{fig:spc}, we show the spectral function $\rho_0( \bm{k},\omega )$ obtained in this approximation near the 2SC-PT at $\mu=400$~MeV and $\varepsilon=0.01$ (left) and $0.2$ (right) with $\varepsilon=(T-T_{\rm c})/T_{\rm c}$.
One sees clear peak structures around
$\omega = \pm k-\mu $ in both figures, which correspond to the quasi-quark and anti-quark excitations, respectively.
The quasi-quark peak has a clear depression around the Fermi energy $ \omega=0$, 
which means that the decay rate of quark excitations is enhanced there.
The depression becomes more remarkable as $\epsilon$ decreases.
This behavior is in contrast to that of the conventional Fermi liquid, in which the lifetime of the quasiparticles becomes longer as $\omega$ approaches the Fermi energy.

\begin{figure}
\begin{center}
\includegraphics[width=13cm]{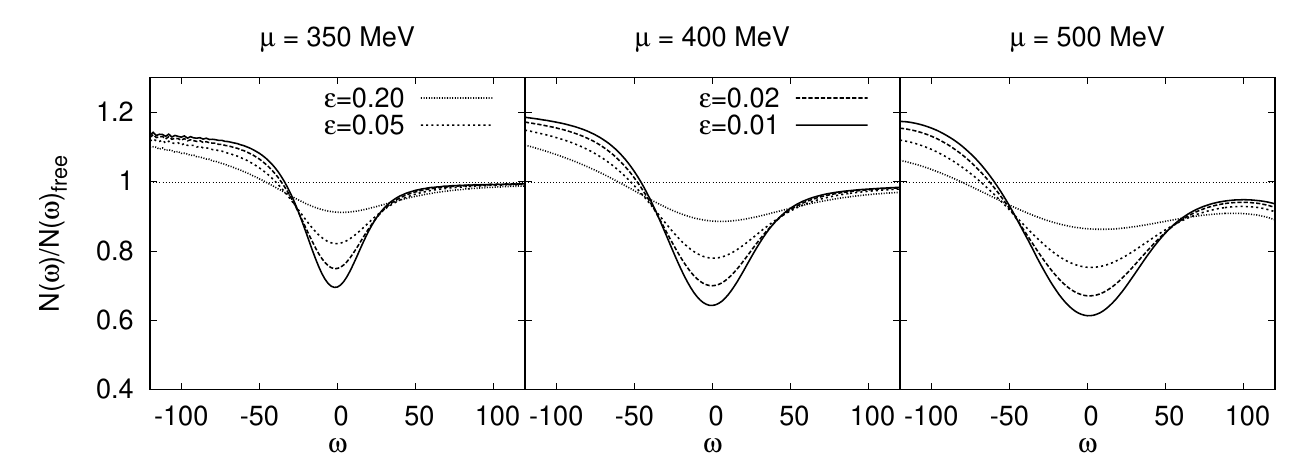} 
\caption{
Density of state at $\mu=400$~MeV and various $\varepsilon\equiv(T-T_{\rm c})/T_{\rm c}$~\cite{Kitazawa:2003cs}.
The Dotted line shows that of the free quarks.
A clear pseudogap structure is seen, which survives
up to $ \varepsilon\approx 0.05$.
}
\label{fig:dos}
\end{center} 
\end{figure}

Substituting this spectral function into Eq.~\eqref{eqn:N_0}, one obtains the quark DOS $N(\omega)$.
In Fig.~\ref{fig:dos}, we show the DOS normalized by that of the free quarks, $N_{\rm free}(\omega)=2N_{\rm f} N_{\rm c} (\omega-\mu)^2/\pi^2$, for $\mu=350,400,500$~MeV and several values of $\epsilon$~\cite{Kitazawa:2003cs}.
There appears clear depression in the DOS around the Fermi energy for $\epsilon=0.01$ for all $\mu$, and the depression survives up to $\epsilon \approx 0.05$.
This result shows the pseudogap phenomenon of the color superconductivity.
The appearance of the pseudogap in the quark DOS is naturally
understood through the non-Fermi liquid behaviors 
in $\rho_0( \bm{k},\omega )$ shown for the first time 
in~\cite{Kitazawa:2003cs,Kitazawa:2005vr}.

Similar pseudogap phenomena owing to the soft modes of the QCD-CP have been explored 
in Ref.~\cite{Kitazawa:2014sga}. In this case, however, it was found that the quark spectrum 
has an intricate rich structure. 
In a thermal relativistic system, it is known that a simple boson-exchange interaction 
leads to the mass gap in fermionic excitations, called the thermal mass. 
When the boson that couples to the fermions is massless, the fermionic excitation 
has two branches, one of which is called the plasmino. 
It is also known that when the boson mass is nonzero and at the same order as $T$, 
the fermionic spectrum exhibits three-peak structure~\cite{Kitazawa:2006zi,Kitazawa:2007ep}; 
in addition to the normal and plasmino modes, there emerges an almost massless mode. 
In Ref.~\cite{Kitazawa:2005mp}, 
the emergence of such a three-peak structure is confirmed near 
the chiral phase transition in the chiral limit at $\mu=0$. 
The analysis is then extended to the QCD-CP in Ref.~\cite{Kitazawa:2013zya,Kitazawa:2014sga}.

\section{Electric conductivity and dilepton production rates}
\label{sec:sigma,DPR}

In this section, we explore the effects of the soft modes on the electric conductivity and dilepton production rates (DPR) near the 2SC-PT and QCD-CP. These quantities are derived from the 
retarded photon self-energy 
\begin{align}
    \Pi^{R\mu\nu}(\bm{k},\omega)
    = \int d^4x e^{i\omega t - i \bm{k}\cdot \bm{x}}\langle [ j^\mu(\bm{x}, t ), j^\nu(\bm{0},0)]\rangle \theta(t),
    \label{eq:jj}    
\end{align}
with the electric current operator $j^\mu(\bm{x},t)$.
The electric conductivity $\sigma$ is given by the low-energy limit of Eq.~\eqref{eq:jj} as
\begin{align}
    \sigma = - \frac13 \lim_{\omega\to0} \frac1\omega \sum_{i=1}^3 {\rm Im} \Pi^{Rii}(\bm{0},\omega),
    \label{eq:sigma}
\end{align}
and the DPR is related to $\Pi^{R\mu\nu}(\bm{k},\omega)$ as
\begin{align}
\frac{d^4\Gamma}{d^4k} = - \frac{\alpha}{12\pi^4} 
\frac{1}{k^2} \frac{1}{e^{\omega/T}-1} g_{\mu\nu}  
{\rm Im} \Pi^{R \mu\nu} (k) ,
\label{eq:DPR}
\end{align}
where $\alpha=e^2/4\pi$ is the fine structure constant. 

In this section, 
we first construct the photon self-energy $\Pi^{R\mu\nu}(\bm{k},\omega)$ 
by incorporating the effects of the soft modes so as to satisfy the 
 {\em Ward-Takahashi (WT) identity}, and then derive the electric conductivity and DPR from it.
Throughout this section, we assume that $\Pi^{R\mu\nu}(\bm{k},\omega)$ consists of three parts
\begin{align}
\Pi^{R\mu\nu} (\bm{k},\omega) = \Pi^{R\mu\nu}_{\rm free} (\bm{k},\omega) +
\Pi^{R\mu\nu}_D (\bm{k},\omega) + \Pi^{R\mu\nu}_S (\bm{k},\omega),
\label{eq:Pi-tot}
\end{align}
where $\Pi^{R\mu\nu}_D (\bm{k},\omega)$ and $\Pi^{R\mu\nu}_S (\bm{k},\omega)$ 
represent the contributions from the soft modes of the 2SC-PT and QCD-CP, respectively, 
that will be defined below and $\Pi^{R\mu\nu}_{\rm free} (\bm{k},\omega)$ 
is the self-energy of the free-quark system. In the Matsubara formalism, it is given by
\begin{align}
\tilde\Pi^{\mu\nu}_{\rm free} (k) = 
N_c C_{\rm em} \int_p {\rm Tr}_D
[\gamma^\mu \mathcal{G}_0 (p+k) \gamma^\nu \mathcal{G}_0 (p)],
\label{eq:Pi-free}
\end{align}
with $C_{\rm em} \equiv e_u^2 + e_d^2$, where $e_u = 2|e|/3$  ($e_d = -|e|/3$) is the electric charge of the up (down) quark with 
$e$ being the electron charge.

\subsection{Photon self-energy}
\label{sec:Pi}

\subsubsection{Contribution of the soft modes of 2SC-PT}
\label{sec:Pi:2SC}

\begin{figure}[t]
\centering
\includegraphics[width=0.9\textwidth]{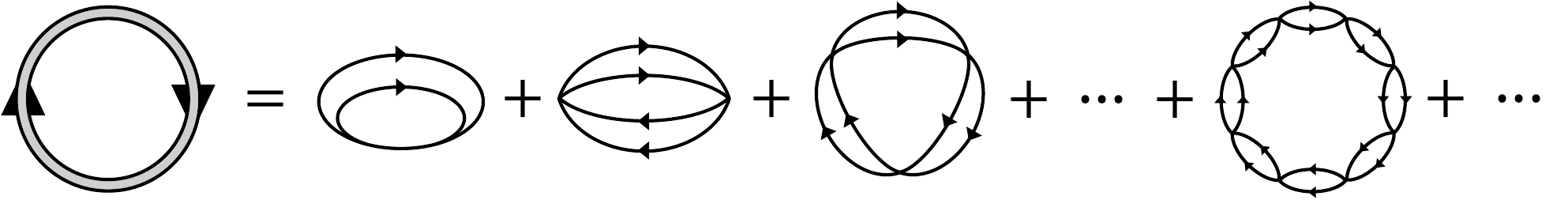}
\caption
{
Contribution of the diquark soft mode to the thermodynamic potential.
}
\label{fig:Omega_2SC}
\end{figure}

\begin{figure}[t]
    \centering
    \begin{tabular}{cccc}
    \includegraphics[width=0.2\textwidth]{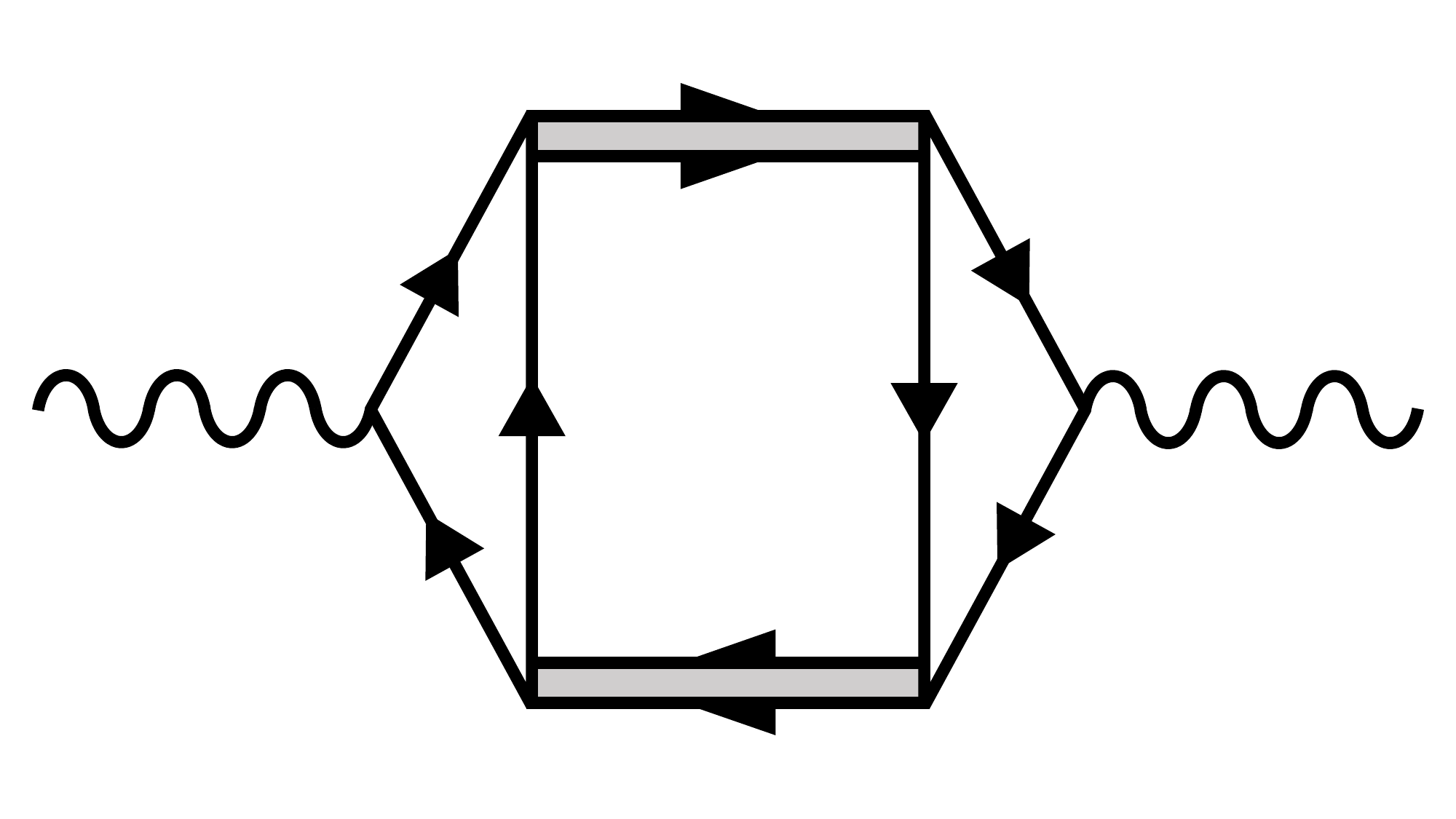} &
    \includegraphics[width=0.2\textwidth]{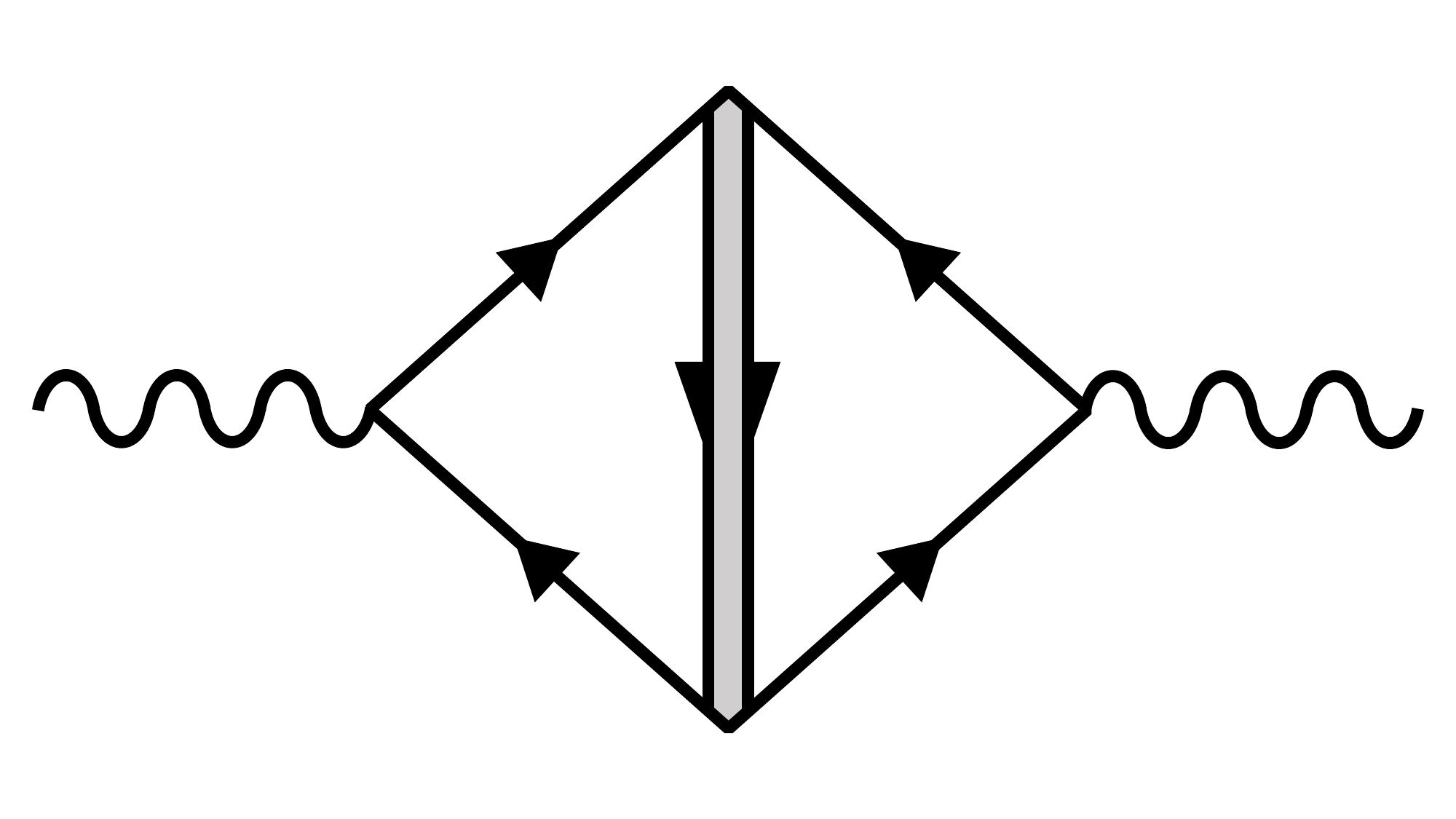} &
    \includegraphics[width=0.2\textwidth]{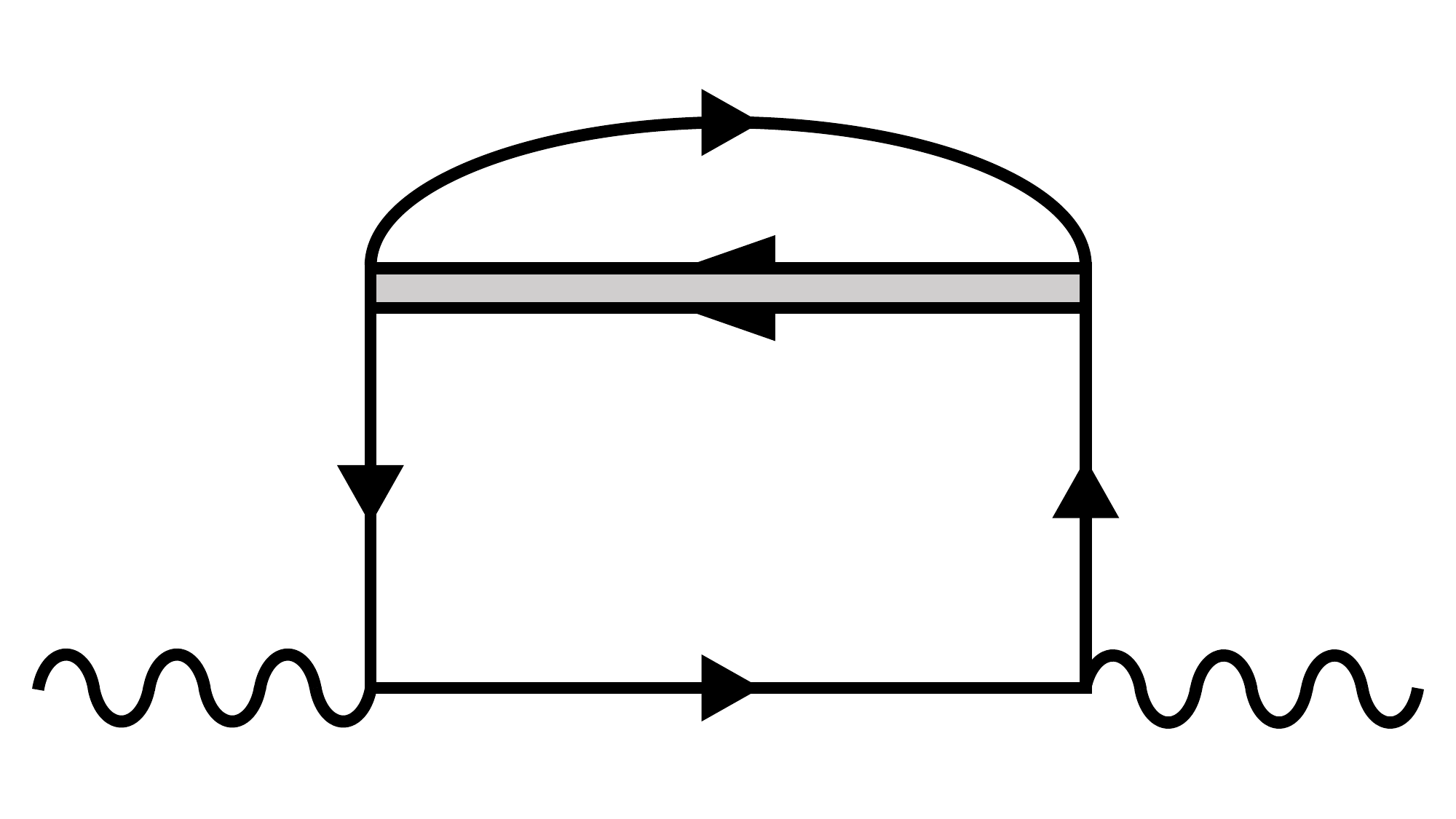} &
    \includegraphics[width=0.2\textwidth]{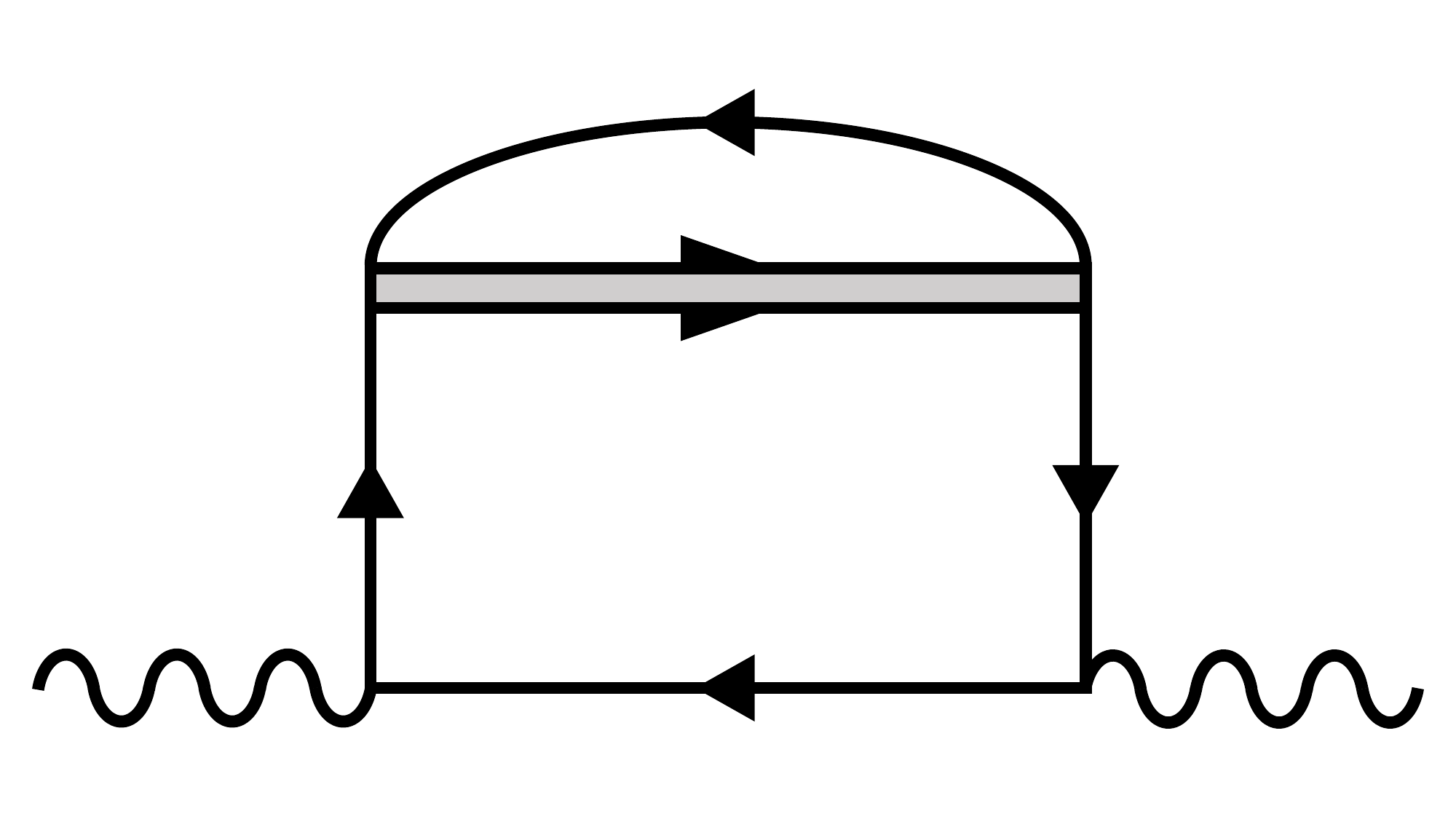}
    \\
    (a) & (b) & (c) & (d)
    \end{tabular}
    \caption{
Diagrammatic representations of the Aslamazov-Larkin (a),
Maki-Thompson (b) and density of states (c, d) terms with the 2SC soft modes
with the wavy lines being the photon ones.
}
\label{fig:Self-energy_2SC}
\end{figure}

Let us first investigate $\Pi^{R\mu\nu}_D(\bm{k},\omega)$, which represents the effects of the soft modes of the 2SC-PT. 
To construct it in a gauge-invariant manner, we start with
 the lowest contribution of the soft modes to the thermodynamic potential 
\begin{align}
\Omega_D = 3 \int_p \ln [G_D \tilde\Xi_D^{-1} (p)], 
\label{eq:Omega_D}
\end{align}
which is the one-loop diagram of $\tilde{\Xi}_D(p)$ given in Fig.~\ref{fig:Omega_2SC}, where $\tilde\Xi_D(p)$
is the imaginary-time $T$-matrix corresponding to Eq.~\eqref{eq:genf-T-matrix} and the overall 
coefficient $3$ comes from 
three anti-symmetric channels of the diquark modes.

The photon self-energy satisfying the WT identity is 
then constructed by attaching electromagnetic vertices 
at any two points on the quark lines in $\Omega_D$.
This procedure leads to the four types of diagrams shown in Fig.~\ref{fig:Self-energy_2SC}; they are called (a) Aslamazov-Larkin (AL)~\cite{AL:1968}, (b) Maki-Thompson (MT)~\cite{Maki:1968,Thompson:1968}, and (c, d) density of states (DOS) terms, respectively, in the theory of metallic superconductivity~\cite{Larkin:book}. 
In the Matsubara formalism we have
\begin{align}
\tilde\Pi^{\mu\nu}_D (k) 
= \tilde\Pi^{\mu\nu}_{{\rm AL},D} (k)
+ \tilde\Pi^{\mu\nu}_{{\rm MT},D} (k) 
+ \tilde\Pi^{\mu\nu}_{{\rm DOS},D} (k) ,
\label{eq:Pi-fluc_2SC}
\end{align}
where 
\begin{align}
\tilde\Pi_{{\rm AL},D}^{\mu\nu} (k) &= 3 \int_q
\tilde\Gamma^\mu_D (q, q+k) \tilde\Xi_D (q+k) 
\tilde\Gamma^\nu_D (q+k, q) \tilde\Xi_D (q),
\label{eq:AL_2SC} \\
\tilde\Pi_{{\rm MT~(DOS)},D}^{\mu\nu} (k) &= 3 \int_q
\tilde\Xi_D (q) \ \mathcal{R}_{{\rm MT~(DOS)}, D}^{\mu\nu} (q, k),
\label{eq:MT_2SC}
\end{align}
denote the contributions of the AL, MT, and DOS terms, respectively, with the vertex functions
\begin{align}
\tilde\Gamma^\mu_D (q, q+k) 
&= 4 (N_c - 1) (e_u+e_d) \int_p
{\rm Tr}_D [\mathcal{G}_0 (p) \gamma^\mu \mathcal{G}_0 (p+k) \mathcal{G}_0 (q-p)],
\label{eq:Gamma_2SC} \\
\mathcal{R}_{{\rm MT},D}^{\mu\nu} (q, k) 
&= 8 (N_c - 1)~e_u e_d \int_p
{\rm Tr}_D [\mathcal{G}_0 (p) \gamma^\mu \mathcal{G}_0 (p+k) \mathcal{G}_0 (q-p-k) 
\gamma^\nu\mathcal{G}_0 (q-p)],
\label{eq:RMT_2SC} \\
\mathcal{R}_{{\rm DOS},D}^{\mu\nu} (q, k) 
&= 4 (N_c - 1) (e_u^2 + e_d^2) \sum_{s=\pm} \int_p
{\rm Tr}_D [\mathcal{G}_0 (p) \gamma^\mu \mathcal{G}_0 (p+sk) \gamma^\nu    
\mathcal{G}_0 (p) \mathcal{G}_0 (q-p)],
\label{eq:RDOS_2SC}
\end{align}
and $q = (\bm{q}, i\nu_n)$ representing the four-momentum of the soft mode. One can explicitly check that they satisfy the WT identities 
\begin{align}
k_\mu \tilde{\Gamma}_D^\mu (q, q+k) 
&= (e_u+e_d) [\mathcal{Q}_D(q+k)-\mathcal{Q}_D(q)], 
\label{eq:AL-vertex_D} 
\\
k_\mu \mathcal{R}_D^{\mu\nu} (q, k) 
&= (e_u+e_d) [\tilde{\Gamma}_D^\nu (q-k, q)-\tilde{\Gamma}_D^\nu (q, q+k)] ,
\label{eq:MTDOS-vertex_D}
\end{align}
with $\mathcal{R}_D^{\mu\nu} (q, k) = \mathcal{R}_{{\rm MT},D}^{\mu\nu} (q, k) + \mathcal{R}_{{\rm DOS},D}^{\mu\nu} (q, k) $.
Using Eqs.~\eqref{eq:AL-vertex_D} and~\eqref{eq:MTDOS-vertex_D}, it is also verified that $\tilde{\Pi}^{\mu\nu} (k)$ satisfies the WT identity $k_\mu\tilde{\Pi}^{\mu\nu} (k)=0$.

In the following numerical analyses, we use the LE or TDGL approximation for $\tilde\Xi_D(q)$ in Eqs.~\eqref{eq:AL_2SC} and~\eqref{eq:MT_2SC}. In this case, the vertices $\tilde{\Gamma}_D^\mu(q, q+k)$ and  
$\mathcal{R}^{\mu\nu}_D(q, k)$ are determined so as to satisfy Eqs.~\eqref{eq:AL-vertex_D} and~\eqref{eq:MTDOS-vertex_D} with these approximated $T$-matrices. This procedure is most easily carried out with the low energy-momentum expansion of the vertices~\cite{Nishimura:2024kvz}.

When the LE approximation is adopted, one can show that $\mathcal{R}_D^{\mu\nu} (q, k)$ is a real function of momentum only. Using this property, it is shown that 
\begin{align}
    {\rm Im}\big[\Pi^{Rij}_{{\rm MT},D} (\bm{k}, \omega) 
              + \Pi^{Rij}_{{\rm DOS},D} (\bm{k}, \omega) \big] = 0 ,
\label{eq:MT-DOS_cancel_2SC}
\end{align}
i.e., the MT and DOS terms cancel out exactly in ${\rm Im}\Pi^{Rij}(\bm{k},\omega)$~\cite{Nishimura:2023not}.
Since the electric conductivity and the DPR depend only on ${\rm Im}\Pi^{Rij}(\bm{k},\omega)$ as in Eqs.~\eqref{eq:sigma} and~\eqref{eq:DPR}, Eq.~\eqref{eq:MT-DOS_cancel_2SC} shows that only the AL term contributes to these quantities.
The explicit form of ${\rm Im} \Pi^{Rij}_D (\bm{k},\omega)$
is obtained from Eq.~\eqref{eq:AL_2SC} by taking the analytic continuation $i\nu_l \rightarrow \omega + i\eta$ as
\begin{align}
{\rm Im} \Pi^{R ij}_D (\bm{k}, \omega) 
=&\ {\rm Im} \Pi^{R ij}_{{\rm AL}, D} (\bm{k}, \omega) \nonumber \\
=&~3 \int \frac{d^3q}{(2\pi)^3}
\Gamma_D^i (\bm{q}, \bm{q}+\bm{k}) \Gamma_D^j (\bm{q}+\bm{k}, \bm{q})
\int \frac{d\omega'}{2\pi} {\rm coth} \frac{\omega'}{2T}
 \nonumber \\
&\ {\rm Im} \Xi^R_D (\bm{q}, \omega')
\Big\{ {\rm Im} \Xi^R_D (\bm{q} + \bm{k}, \omega'+\omega) 
- {\rm Im} \Xi^R_D (\bm{q} - \bm{k}, \omega'-\omega) \Big\}.
\label{eq:AL-explicit_2SC}
\end{align}

\subsubsection{Contribution of the soft modes of QCD-CP}
\label{sec:Pi:QCDCP}

Next, we calculate $\Pi_S^{R\mu\nu}(\bm{k},\omega)$, i.e., the modification of the photon
 self-energy from the soft modes of the QCD-CP~\cite{Nishimura:2023oqn}.
As the calculational procedure goes in a similar way to Sec.~\ref{sec:Pi:2SC}, duplicated descriptions will be omitted in what follows.

\begin{figure}[t]
\centering
\includegraphics[width=0.9\textwidth]{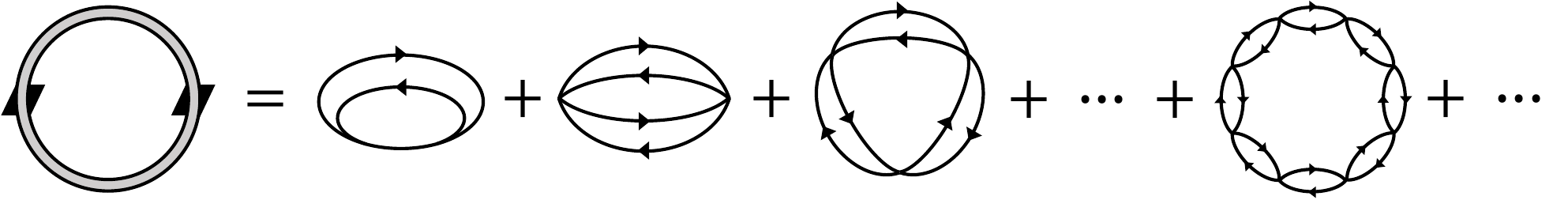}
\caption
{
Contribution of the soft mode of the QCD-CP to the thermodynamic potential.
}
\label{fig:Omega_QCDCP}
\end{figure}

\begin{figure*}[t]
\centering

\begin{tabular}{c@{\hspace{1pt}}c@{\hspace{1pt}}c@{\hspace{1pt}}c@{\hspace{1pt}}c}
        \includegraphics[width=0.18\textwidth]{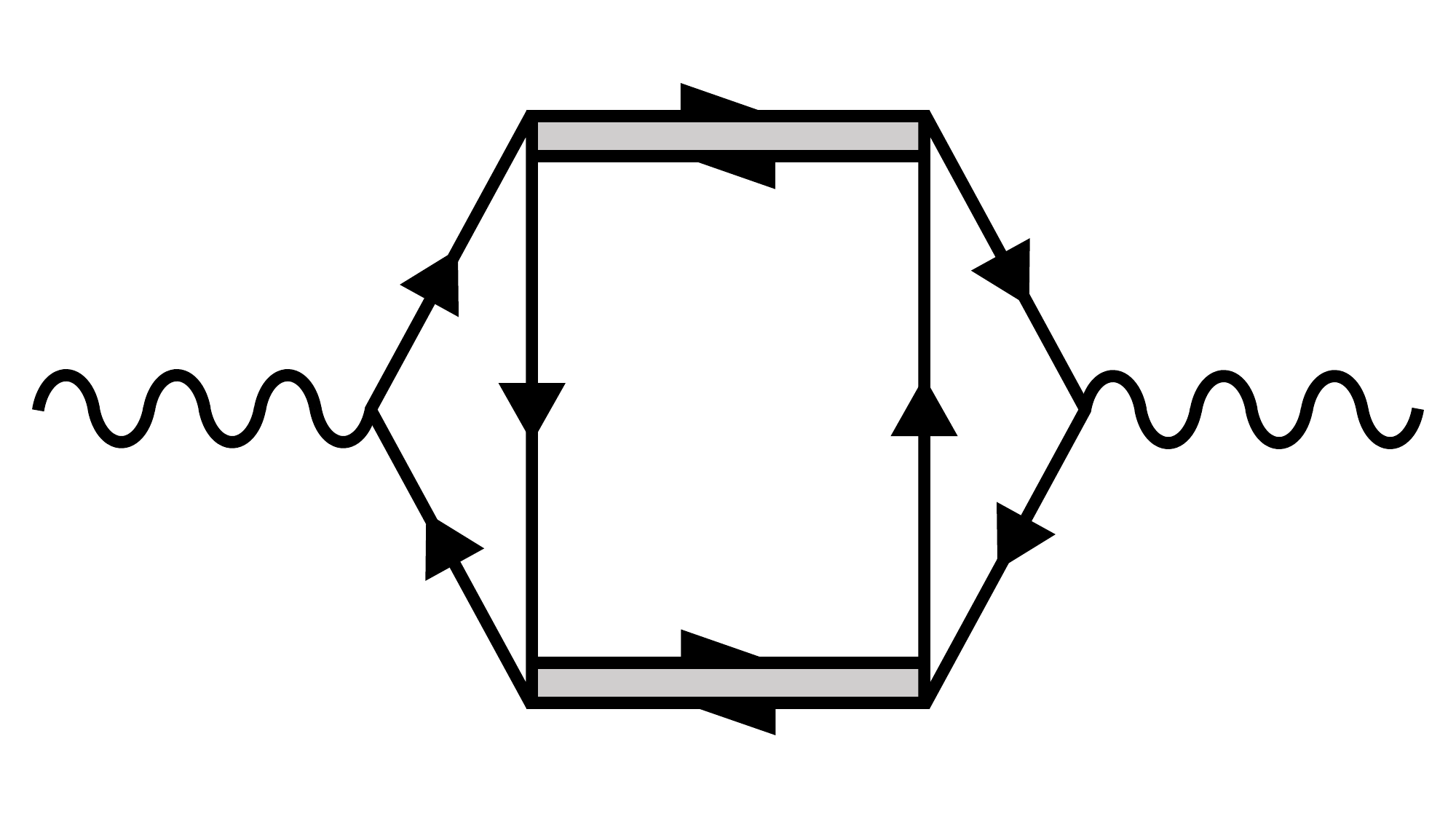} &
        \includegraphics[width=0.18\textwidth]{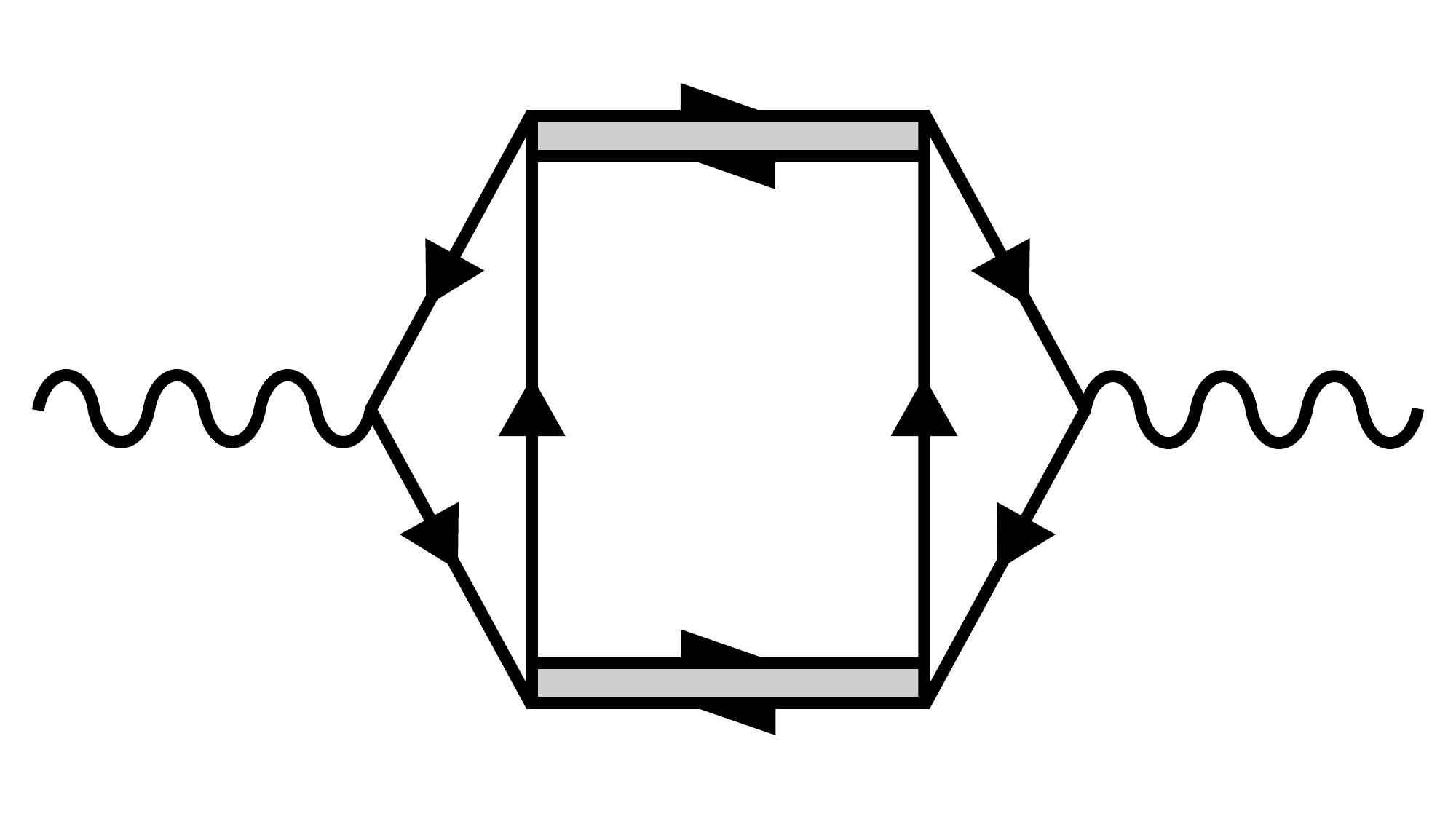} &
        \includegraphics[width=0.18\textwidth]{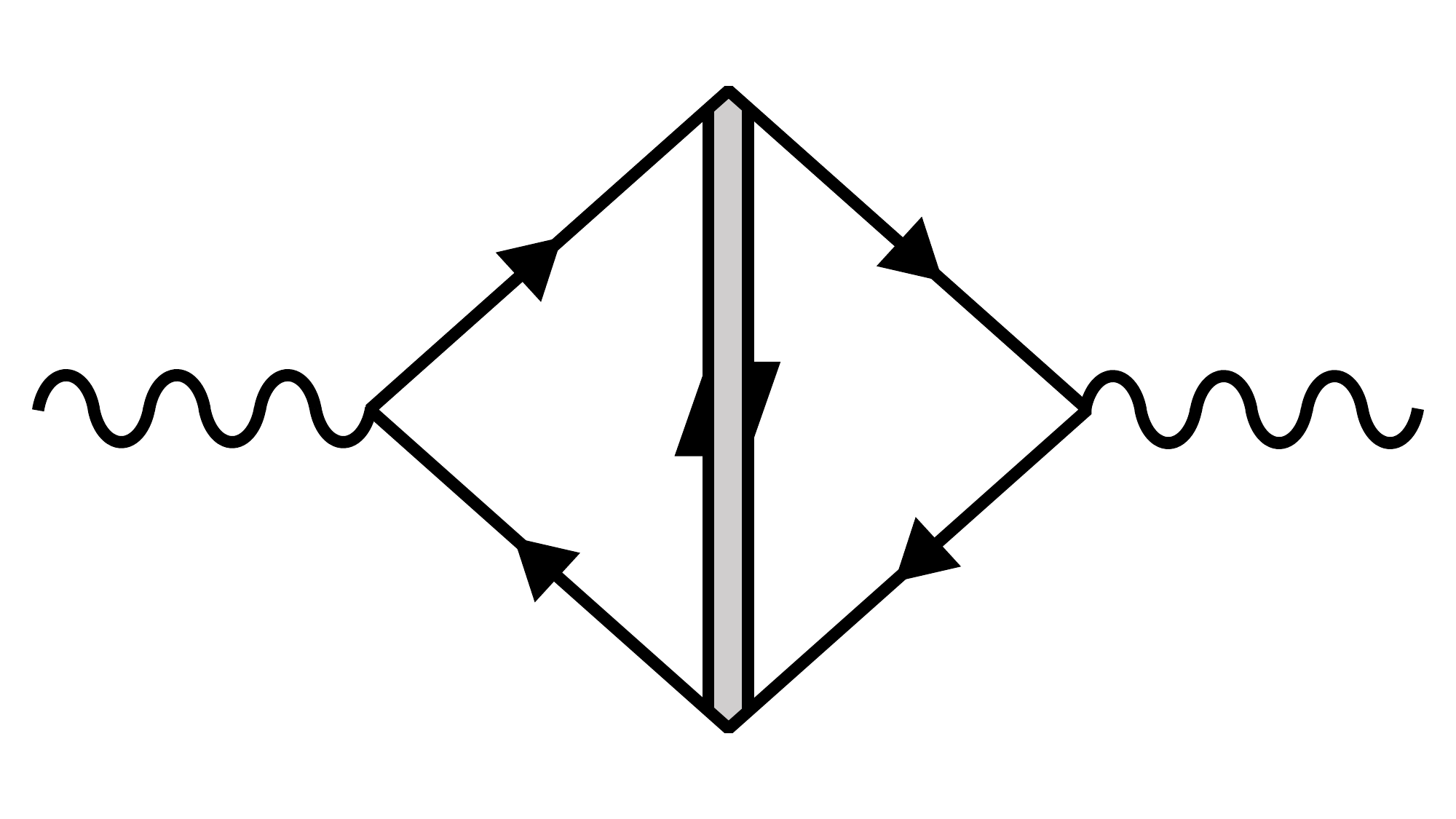} &
        \includegraphics[width=0.18\textwidth]{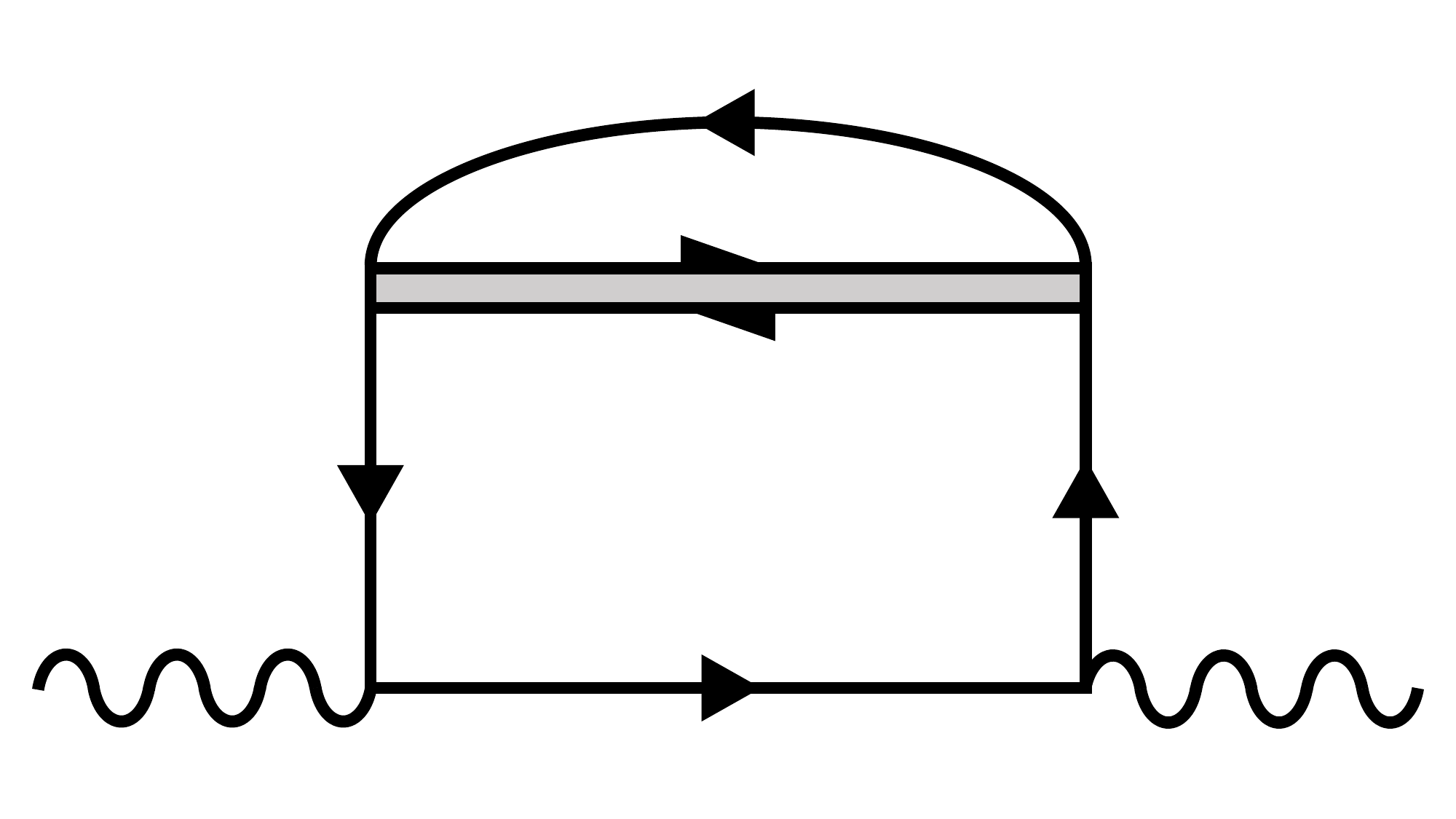} &
        \includegraphics[width=0.18\textwidth]{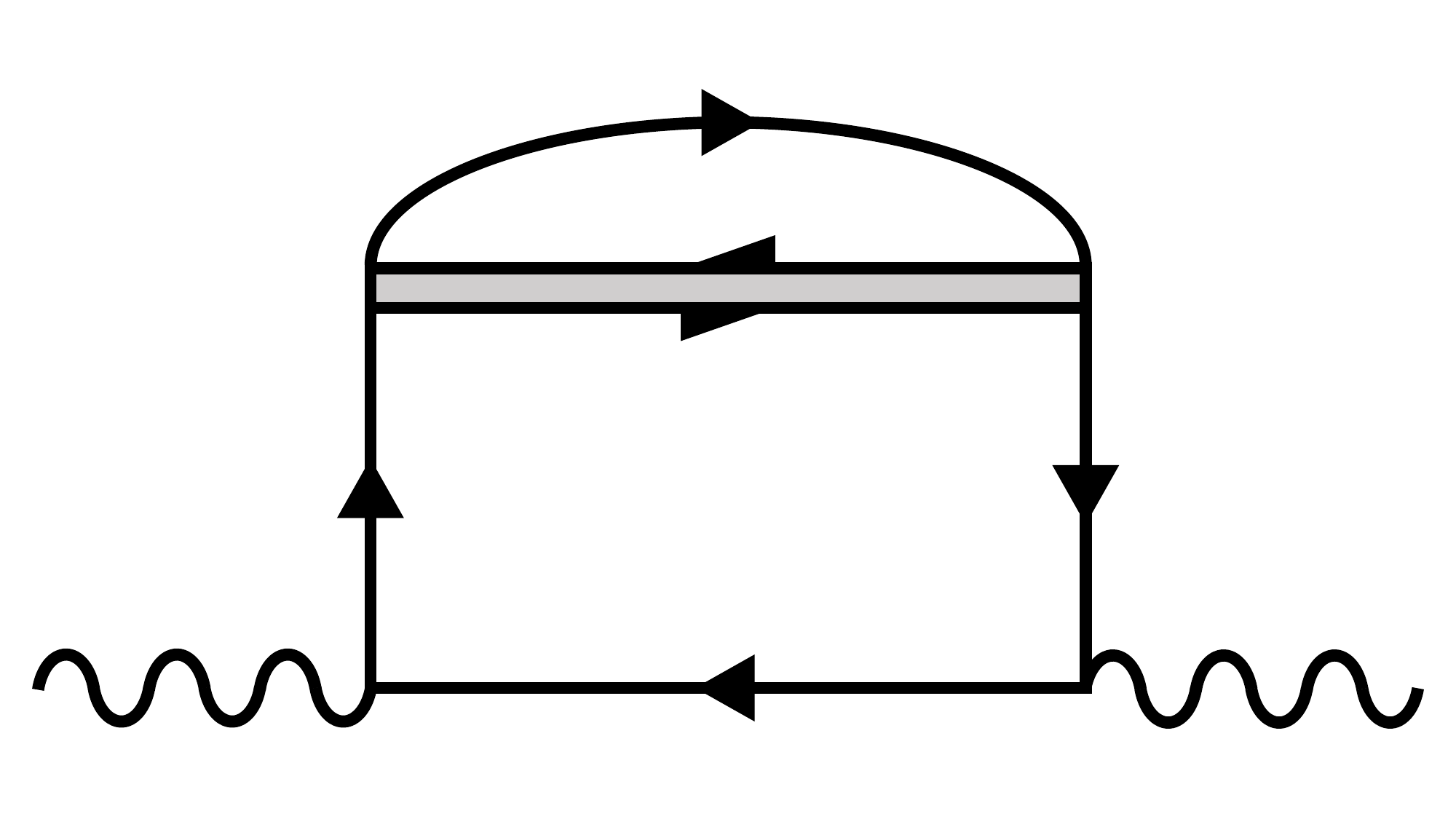}
        \\
        (a) & (c) & (e) & (g) & (i)
        \\
        \\
        \includegraphics[width=0.18\textwidth]{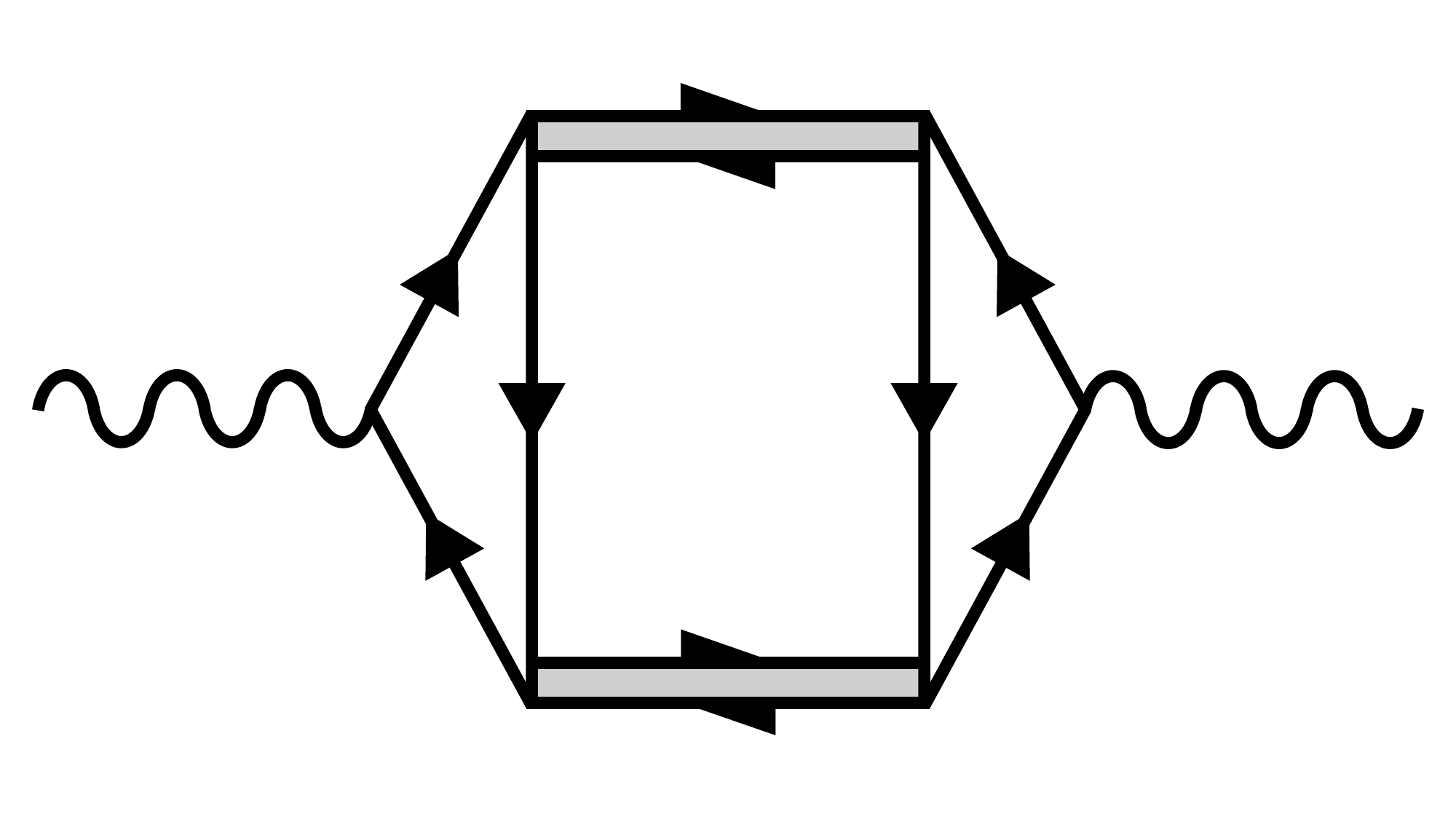} &
        \includegraphics[width=0.18\textwidth]{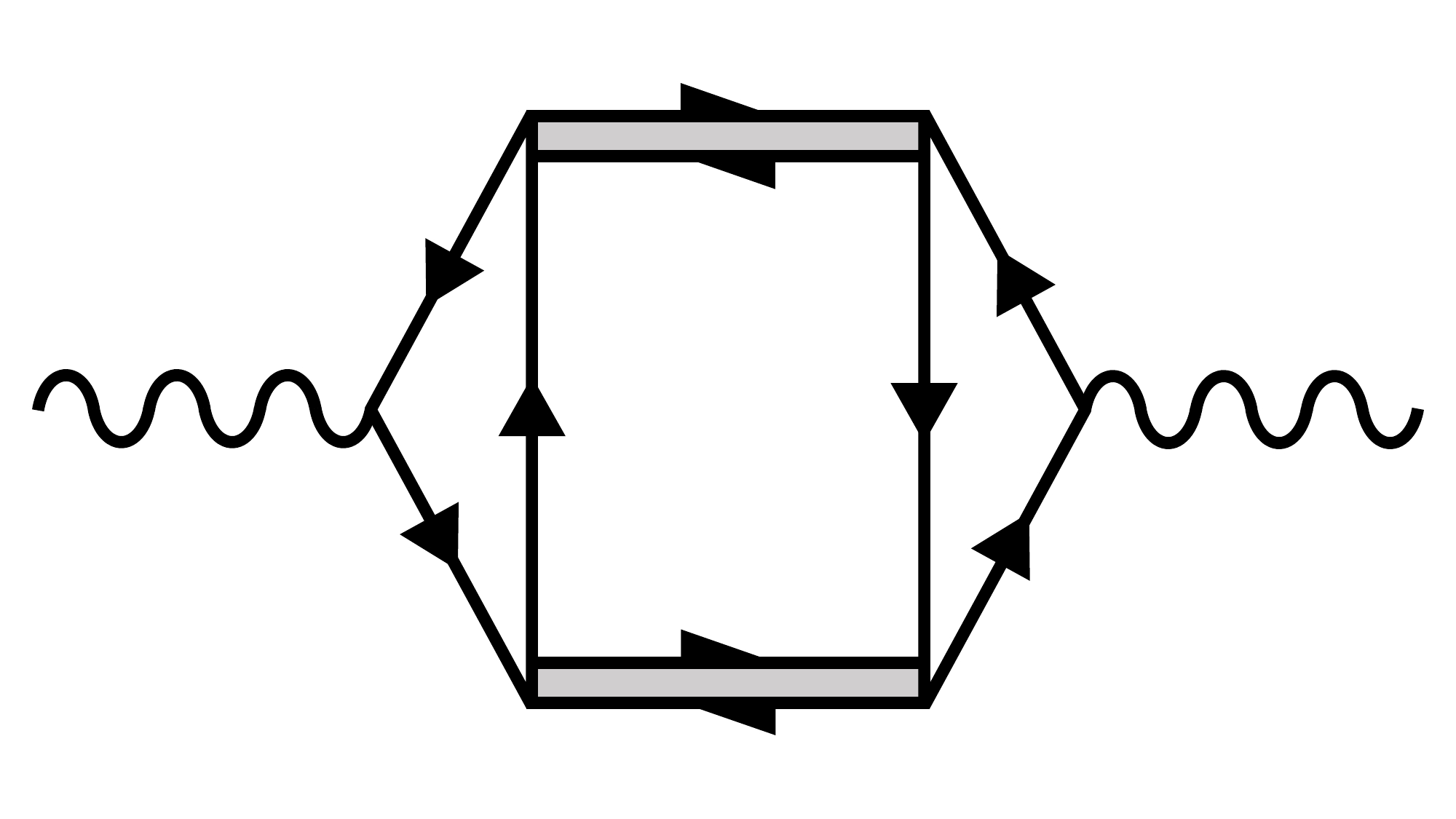} &
        \includegraphics[width=0.18\textwidth]{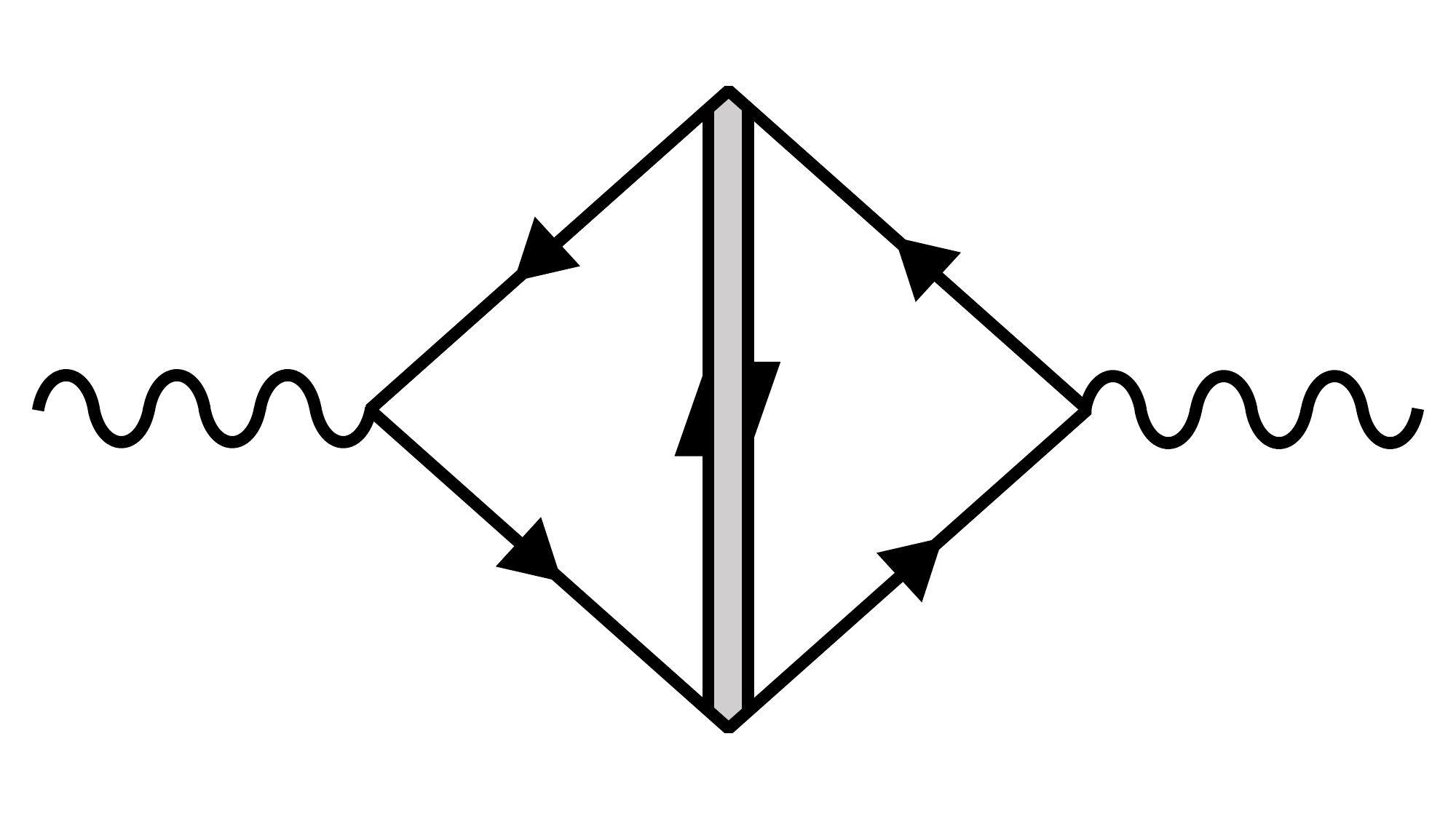} &
        \includegraphics[width=0.18\textwidth]{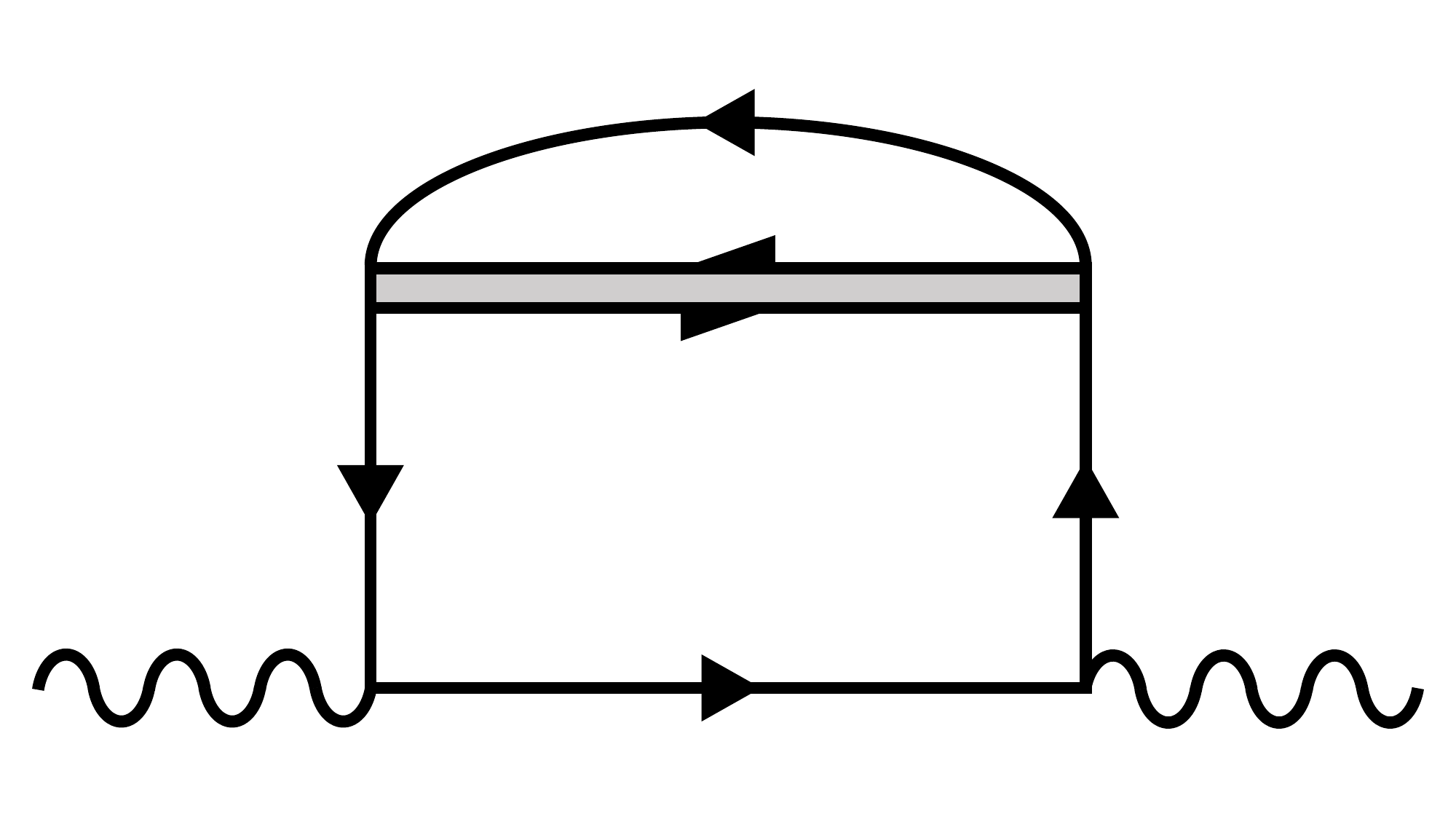} &
        \includegraphics[width=0.18\textwidth]{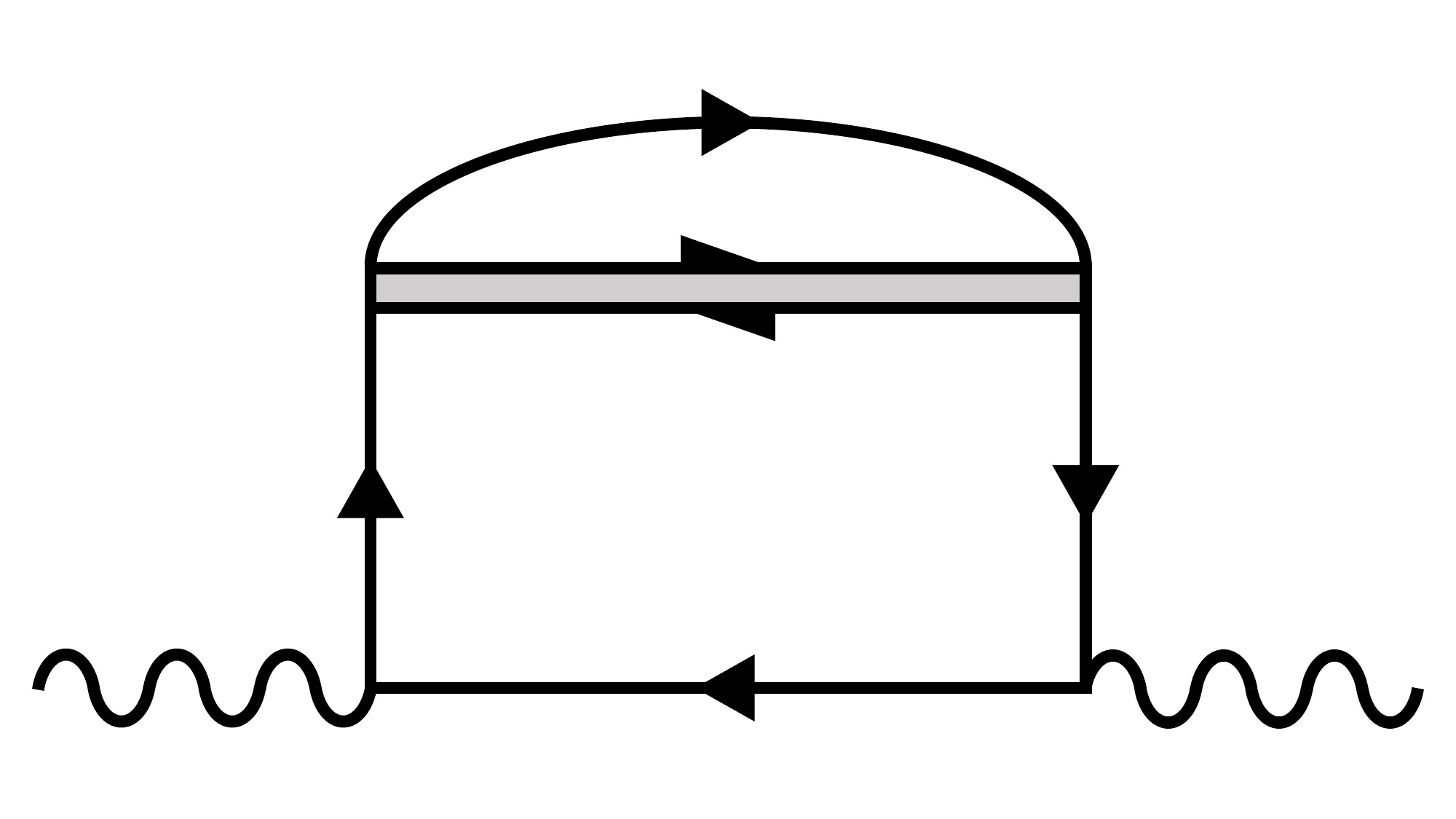}
        \\
        (b) & (d) & (f) & (h) & (j)
\end{tabular}
    \caption{
Diagrammatic representations of the Aslamazov-Larkin (a)--(d),
Maki-Thompson (e, f) and density of states (g)--(j) terms 
with the soft modes of the QCD-CP.
The single, double, and wavy lines are quarks, 
soft modes, and photon, respectively.
}
\label{fig:Self-energy_QCDCP}
\end{figure*}

Similarly to the previous analysis, we start from the one-loop diagram of the soft mode of the QCD-CP shown in Fig.~\ref{fig:Omega_QCDCP},
which is the lowest-order contribution to the thermodynamic potential
$ \Omega_S = \int_p \ln [G_S \tilde\Xi_S^{-1}(p)]$.
Attaching electromagnetic vertices at any two different points of quark lines in $\Omega_S$,
we obtain ten types of diagrams shown in Fig.~\ref{fig:Self-energy_QCDCP}, 
where the number of diagrams is 
larger than that in Fig.~\ref{fig:Self-energy_2SC} because the directions of quark lines in the vertices should be distinguished in this case.
We refer to the diagrams (a)--(d) as the AL, (e), (f) as the MT, and (g)--(j) as the DOS terms, respectively.
The respective contributions to the photon self-energy in the imaginary-time formalism are denoted by 
\begin{align}
\tilde\Pi_{{\rm AL},S}^{\mu\nu} (k) &= \sum_{f=u,d} \int_q
\tilde\Gamma^\mu_f (q, q+k) \tilde\Xi_S (q+k) 
\tilde\Gamma^\nu_f (q+k, q) \tilde\Xi_S (q),
\label{eq:AL_QCDCP} \\
\tilde\Pi_{{\rm MT},S}^{\mu\nu} (k) &= \sum_{f=u,d} \int_q
\tilde\Xi_S (q) \ \mathcal{R}_{{\rm MT},f}^{\mu\nu} (q, k),
\label{eq:MT_QCDCP} \\
\tilde\Pi_{{\rm DOS},S}^{\mu\nu} (k) &= \sum_{f=u,d} \int_q
\tilde\Xi_S (q) \ \mathcal{R}_{{\rm DOS},f}^{\mu\nu} (q, k),
\label{eq:DOS_QCDCP} 
\end{align}
which in total gives $\tilde\Pi^{\mu\nu}_S (k)$
as 
\begin{align}
\tilde\Pi^{\mu\nu}_S (k) 
= \tilde\Pi^{\mu\nu}_{{\rm AL},S} (k)
+ \tilde\Pi^{\mu\nu}_{{\rm MT},S} (k) 
+ \tilde\Pi^{\mu\nu}_{{\rm DOS},S} (k).
\label{eq:Pi-fluc_QCDCP}
\end{align}
As before, one can explicitly check that the vertex functions in these equations satisfy the WT identities
\begin{align}
k_\mu \tilde{\Gamma}^\mu_f (q, q+k) &
= -e_f [\tilde{\Xi}_S^{-1} (q+k) - \tilde{\Xi}_S^{-1} (q)], 
\label{eq:Gamma-WT_QCDCP} \\
k_\mu \mathcal{R}^{\mu\nu}_f (q, k) &
= -e_f [\tilde{\Gamma}^\nu_f (q-k, q)-\tilde{\Gamma}^\nu_f (q, q+k)], 
\label{eq:R-WT_QCDCP}
\end{align}
with $\mathcal{R}^{\mu\nu}_f (q, k)=
\mathcal{R}_{{\rm MT},f}^{\mu\nu}(q, k) + \mathcal{R}_{{\rm DOS},f}^{\mu\nu}(q, k)$.
The WT identity of Eq.~\eqref{eq:Pi-fluc_QCDCP},
$k_\mu \tilde\Pi^{\mu\nu}_S (k) = 0$,
is easily verified using Eqs.~\eqref{eq:Gamma-WT_QCDCP} and~\eqref{eq:R-WT_QCDCP}.

As in the previous subsection, we use the LE or TDGL approximation for $\tilde\Xi_S(q)$ and determine the vertex functions $\tilde{\Gamma}^\mu_f (q, q+k)$ and $\mathcal{R}_{f}^{\mu\nu} (q, k)$ so that they satisfy Eqs.~\eqref{eq:Gamma-WT_QCDCP} and~\eqref{eq:R-WT_QCDCP}.
One can then show that $\mathcal{R}^{ij}_f (q, k)$ is a real function only of momenta $\bm{q}$ and $\bm{k}$ as before.
Using this property and the same procedure as we did in Eq.~\eqref{eq:MT-DOS_cancel_2SC}, it is shown that the MT and DOS terms cancel out in the spatial components of 
${\rm Im}\Pi_S^{R\mu\nu}(\bm{k},\omega)$. 
Thus ${\rm Im} \Pi^{R ij}_S (\bm{k}, \omega)$ is again given solely by the AL term.
The final result is calculated to be~\cite{Nishimura:2024kvz}
\begin{align}
{\rm Im} \Pi^{R ij}_S (\bm{k}, \omega)
=&~ {\rm Im} \Pi^{R ij}_{{\rm AL}, S} (\bm{k}, \omega) \nonumber \\
=&~ \sum_f \int \frac{d^3q}{(2\pi)^3}
\Gamma_f^i (\bm{q}, \bm{q}+\bm{k}) \Gamma_f^j (\bm{q}+\bm{k}, \bm{q}) \int \frac{d\omega'}{2\pi} \coth \frac{\omega'}{2T}
\nonumber \\
\times&~ {\rm Im} \Xi^R_S (\bm{q}, \omega')
\Big\{ {\rm Im} \Xi^R_S (\bm{q} + \bm{k}, \omega'+\omega) 
- {\rm Im} \Xi^R_S (\bm{q} - \bm{k}, \omega'-\omega) \Big\}.
\label{eq:AL-explicit_QCDCP}
\end{align}

\subsection{Electric conductivity}
\label{sec:Transport-coeff}

Now, we apply the photon self-energy obtained above to the analysis of the electric conductivity $\sigma$. Since the photon self-energy consists of three contributions as given in Eq.~\eqref{eq:Pi-tot},
the spectral function at zero momentum $\rho(\omega)=-\sum_i{\rm Im} \Pi^{Rii}(\bm{0},\omega)$ is also decomposed as 
\begin{align}
    \rho(\omega) = \rho_{\rm free}(\omega) + \rho_D(\omega) + \rho_S(\omega),
\end{align}
with 
$\rho_{\rm free}(\omega)=-\sum_i {\rm Im} \Pi_{\rm free}^{R ii} (\bm{0},\omega)$,
and 
$\rho_\gamma (\omega) = -\sum_i {\rm Im} \Pi_\gamma^{R ii} (\bm{0}, \omega)$ ($\gamma = D,\, S$).
Among them, $\rho_{\rm free}(\omega)$ does not contribute to $\sigma$ since $\rho_{\rm free}(\omega)=0$ for $|\omega|<2M$.
Near the 2SC-PT, $\rho_D(\omega)$ dominates over $\rho_S(\omega)$ and 
the behavior of $\sigma$ is described solely by $\rho_D(\omega)$, and vice versa.

Before discussing the numerical results, it is instructive to explore the behaviors of $\sigma$ near the 2SC-PT and QCD-CP analytically. 
From Eqs.~\eqref{eq:AL-explicit_2SC} and~\eqref{eq:AL-explicit_QCDCP}, derivatives 
of $\rho_\gamma(\omega)$ at $\omega=0$ are obtained as 
\begin{align}
\frac{\partial^n \rho_\gamma(\omega)}{\partial^n \omega} \bigg|_{\omega=0}
&= ~2\bar{N}_\gamma
\int \frac{d^3q}{(2\pi)^3} ~|\bm\Gamma_\gamma(\bm{q},\bm{q})|^2
\int d\omega' \coth\frac{\omega'}{2T} 
~{\rm Im}\Xi^R_\gamma(\bm{q}, \omega')
~\frac{\partial^n}{\partial^n \omega'} {\rm Im}\Xi_\gamma^R(\bm{q}, \omega') ,
\label{eq:rho_gamma_explicit}
\end{align}
with
$\bar{N}_D = 3(e_u+e_d)^2$, 
$|\bm\Gamma_D(\bm{q},\bm{q})|^2 = \sum_i \Gamma_D^i(\bm{q},\bm{q})^2$,
$\bar{N}_S = e_u^2+e_d^2$, and
$|\bm\Gamma_S(\bm{q},\bm{q})|^2 = \sum_{i,f} \Gamma_f^i(\bm{q},\bm{q})^2$.

In the case of the 2SC-PT,
as $T$ approaches $T_{\rm c}$ the integrand in Eq.~\eqref{eq:rho_gamma_explicit} 
diverges at $(|\bm{q}|,\omega')=(0,0)$ owing to $a_D\to0$ in this limit. 
Hence, the dominant contribution to Eq.~\eqref{eq:rho_gamma_explicit}  comes from the origin. 
This justifies the use of the TDGL approximations~\eqref{eq:Xi-TDGL_2SC} and one obtains 
\begin{align}
\sigma = -\frac{3 e_\Delta^2 T}{16 \pi} 
\frac{1}{a_D^{1/2} b_D^{1/2}}
\frac{|c_D|^2}{{\rm Im} c_D} 
\sim T \epsilon^{-1/2} .
\label{eq:coefficients-behavior_2SC}
\end{align}
This result shows that $\sigma$ diverges at $T = T_{\rm c}$ with 
the critical exponent $-1/2$, which corresponds to the mean-field value.
Equation~\eqref{eq:coefficients-behavior_2SC} also 
tells us 
that the magnitude of $\sigma$ does not have any explicit dependence 
on $\mu$ nor $G_D$, implying that $\sigma$ is insensitive to $\mu$ around the 2SC-PT.

Next, we consider the case of the QCD-CP. In this case, the use of the TDGL approximation~\eqref{eq:Xi-TDGL_QCDCP} leads to 
\begin{align}
\frac{\partial \rho_S (\omega)}{\partial \omega} \bigg|_{\omega=0}
=&~\frac{(e_u^2+e_d^2) T}{2 \pi^3} 
\frac{{\rm Im} c_S}{a_S} {\rm tan}^{-1} \frac{{\rm Im} c_S}{a_S} .
\label{eq:deff-rho-1_QCDCP} 
\end{align}
Using Eq.~\eqref{eq:a_S}, we then obtain asymptotic behaviors 
\begin{align}
\sigma \sim \frac{1}{a_S} &\sim
\begin{cases}
~\epsilon_{\rm CP}^{-1} &\text{along the first-order PT or crossover transition lines,} \\
~\epsilon_{\rm CP}^{-2/3} & \text{otherwise.} 
\end{cases}
\label{eq:sigma-behavior_QCDCP} 
\end{align}
We remark that a similar analysis for the anomalous behavior of the relaxation 
times is made in Ref.~\cite{Nishimura:2024kvz}.

\begin{figure*}[t]
\centering
\includegraphics[width=0.87\textwidth]{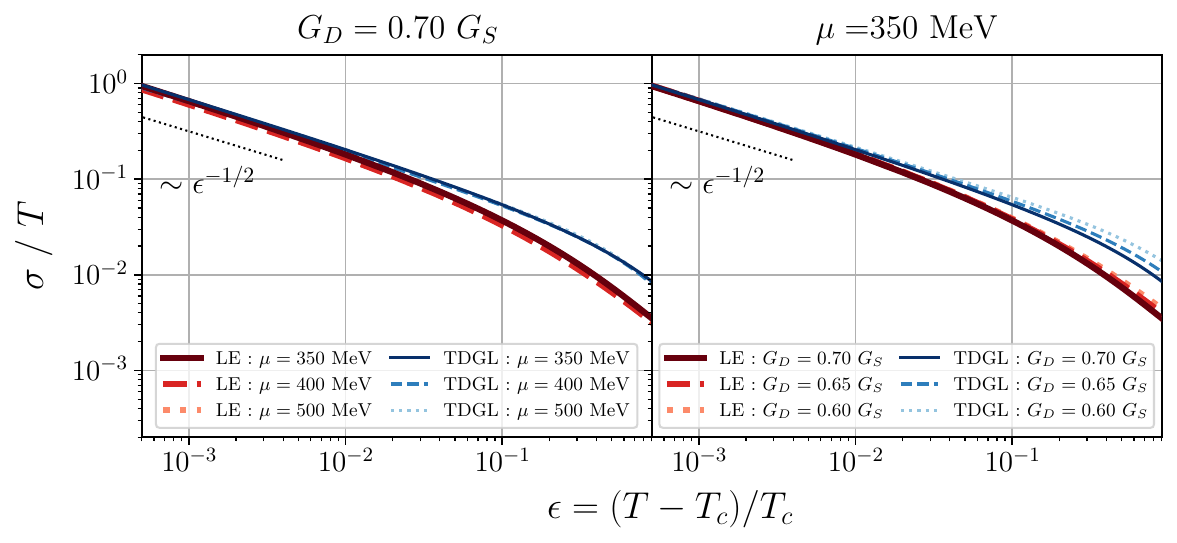}
\caption{
Electric conductivity $\sigma$ 
near the 2SC-PT for 
several values of $\mu$ and $G_D$.
The thick-red and thin-blue lines are the results of the LE and TDGL approximations, respectively.
In the left panels, the lines are plotted 
at $\mu = 350$, $400$, and $500$~MeV with fixed $G_D / G_S= 0.7$,
while the right panels show the results at $G_D / G_S = 0.70$, $0.65$, and $0.60$ for $\mu = 350$~MeV.
The dotted lines indicate the critical exponents $\epsilon^{-1/2}$.
}
\label{fig:sigmaD}
\end{figure*}

\begin{figure*}[t]
\includegraphics[width=0.87\textwidth]{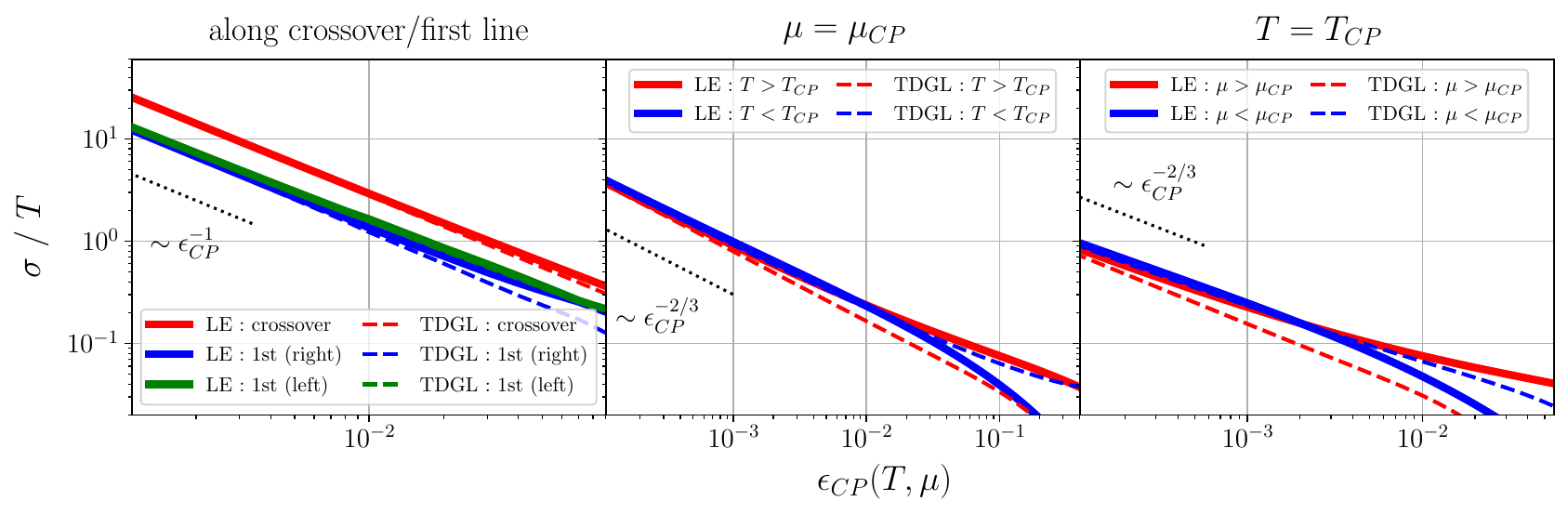}
\caption{
Electric conductivity $\sigma$
near the QCD-CP.
In the left panels, $T$ and $\mu$ are varied along the transition line. 
The middle and right panels show the results with fixed $\mu = \mu_{\rm CP}$ 
and $T = T_{\rm CP}$, respectively.
The dotted line in each panel represents the critical exponents in 
Eqs.~\eqref{eq:sigma-behavior_QCDCP}.
}
\label{fig:sigmaS}
\end{figure*}

Next, let us examine the behavior of $\sigma$ near the 2SC-PT and QCD-CP numerically. 
Figure~\ref{fig:sigmaD} show $\sigma/T$ as a function
 of $\epsilon=(T-T_{\rm c})/T_{\rm c}$.
The left panel shows the results for $\mu=350,400,500$~MeV with fixed $G_D/G_S=0.7$, 
while  the value of $G_D$ is varied at $\mu=350$~MeV in the right panel.
The thick-red (thin-blue) lines show the results 
in the LE (TDGL) approximation:
One can see that $\sigma / T$ grows 
as $T$ approaches $T_{\rm c}$ with the exponents 
in Eq.~\eqref{eq:coefficients-behavior_2SC} indicated by the dotted lines in the figure.
One also sees that the LE and TDGL results tend to approach each other 
in this limit, while their difference grows as $\epsilon$ becomes larger.
The figure confirms that $\sigma/T$ is insensitive to $\mu$ and $G_D$, 
in accordance with the analytical results in Eq.~\eqref{eq:coefficients-behavior_2SC}.

In Fig.~\ref{fig:sigmaS}, 
we show the numerical results for the QCD-CP 
as functions of $\epsilon_{\rm CP}$~\cite{Nishimura:2024kvz}.
In the left panel, $T$ and $\mu$ are varied along the phase transition line.
For the first-order transition side, the results on the transition line are shown 
for the two coexisting states. 
For the crossover side with $T>T_{\rm CP}$, the transition line is defined by the point at which 
the chiral susceptibility,
$\chi_M=\partial^2\Omega/\partial M^2$,
takes the maximum for a given temperature $T$.
In the middle panel, we set $\mu=\mu_{\rm CP}$ and vary $T$, while in the right panel
 $\mu$ is varied with fixed $T=T_{\rm CP}$.
The thick and thin lines are the result of
the LE and TDGL approximations, respectively.
The thin-dotted lines indicate the critical exponents in Eqs.~\eqref{eq:sigma-behavior_QCDCP}. 
We see  that the numerical results are in good agreement 
with the analytical results near the QCD-CP.

\begin{figure}[t]
\centering
\includegraphics[width=0.8\textwidth]{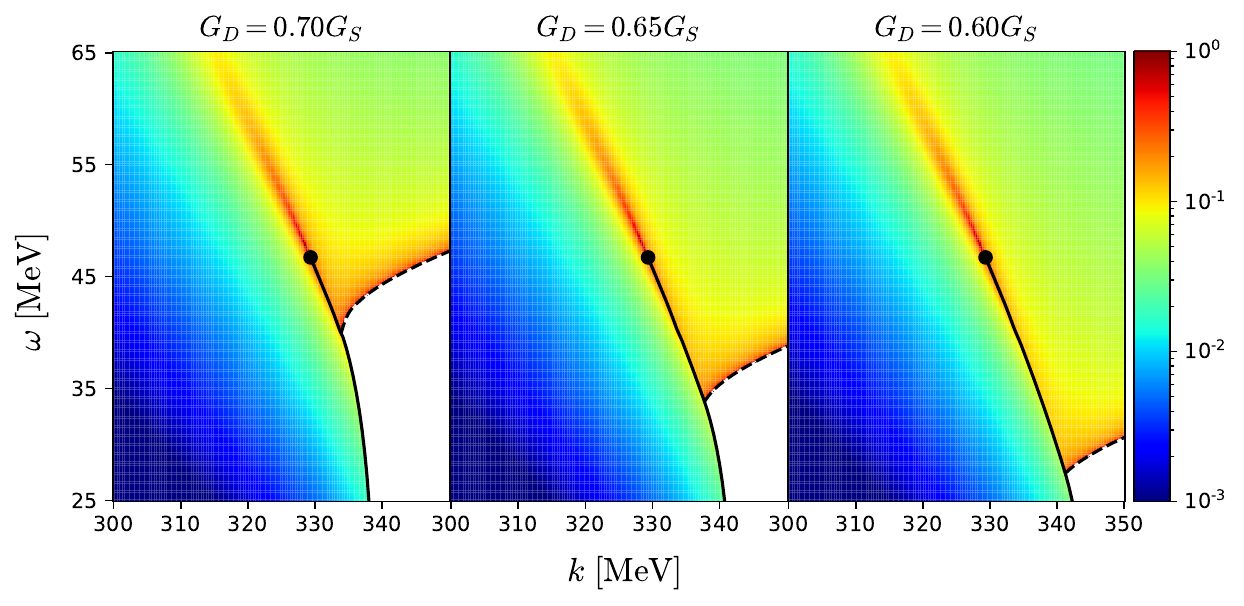}
\caption{
Contour maps of $\sigma/T$ on the $T$--$\mu$ plane 
around the CP with $G_D / G_S= 0.70$, $0.65$ and $0.60$.
The solid and dashed lines are the first-order and second-order phase transitions, respectively.
}
\label{fig:sigma_3d}
\end{figure}

Figure~\ref{fig:sigma_3d} shows a summarizing global behavior of 
$\sigma / T$ as a function of $T$ and $\mu$ when
the effects of the QCD-CP and 2SC-PT are included simultaneously~\cite{Nishimura:2024kvz}.
The color maps show the results 
for three values of the diquark couplings $G_D / G_S= 0.70$, $0.65$ and $0.60$, respectively, 
 in the LE approximation.
The solid and dashed lines denote the first-order transition and the second-order 2SC-PT, 
respectively.
Since our formalism is not applicable to the 2SC phase with $\Delta\ne0$, 
this phase is left blank in the figure.
One finds that $\sigma/T$ is enhanced around the QCD-CP and 2SC-PT.
A significant enhancement due to the presence of the QCD-CP occurs along the critical line
parallel to the first-order transition line.
The existence of two isolated regions of the enhancement of $\sigma$ is interesting 
in light of the beam-energy scan in the HIC, as it would result 
in two non-monotonic behaviors of an experimental observable 
as a function of the collision energy. We will come back to this point in the next subsection.

\subsection{Dilepton production rates}
\label{sec:DPR}

\begin{figure}[tbp]
\centering
\includegraphics[keepaspectratio, scale=0.45]{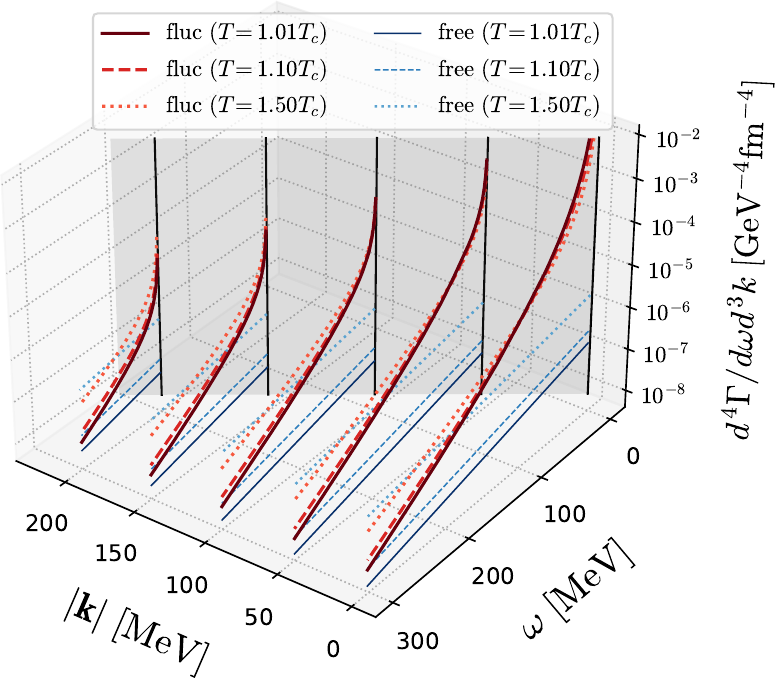}
\includegraphics[keepaspectratio, scale=0.45]{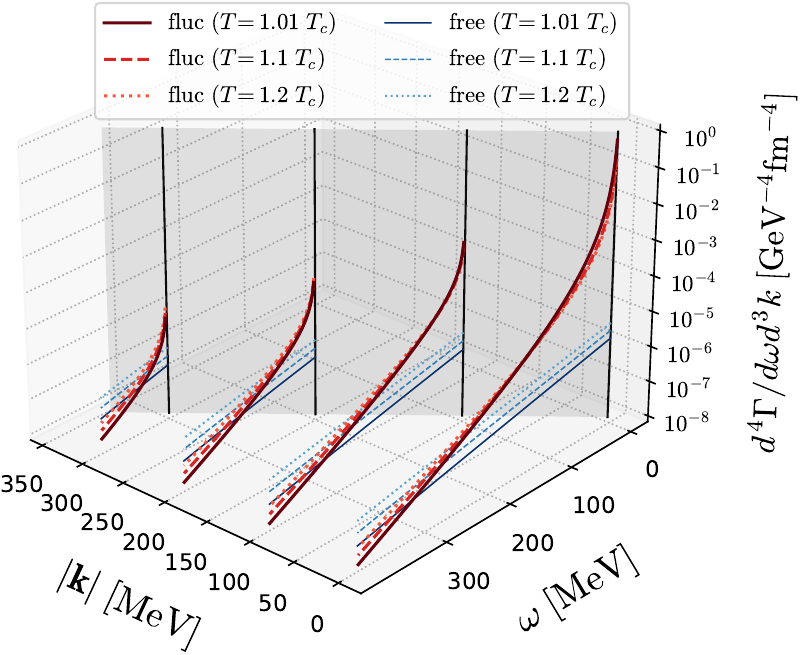}
\caption{
Dilepton production rates per unit energy $\omega$ and momentum $\bm{k}$ above $T_c$ of the 2SC at $\mu = 350~{\rm MeV}$ (left)~\cite{Nishimura:2022mku} and of the QCD-CP at $\mu = \mu_{\rm CP}$ (right)~\cite{Nishimura:2023oqn} with $G_D = 0.7G_S$. 
The thick (thin) lines are the contribution of the soft modes (the massless free quark gases). 
}
\label{fig:DPR}
\end{figure}

\begin{figure}[tbp]
\centering
\includegraphics[width=0.45\linewidth]{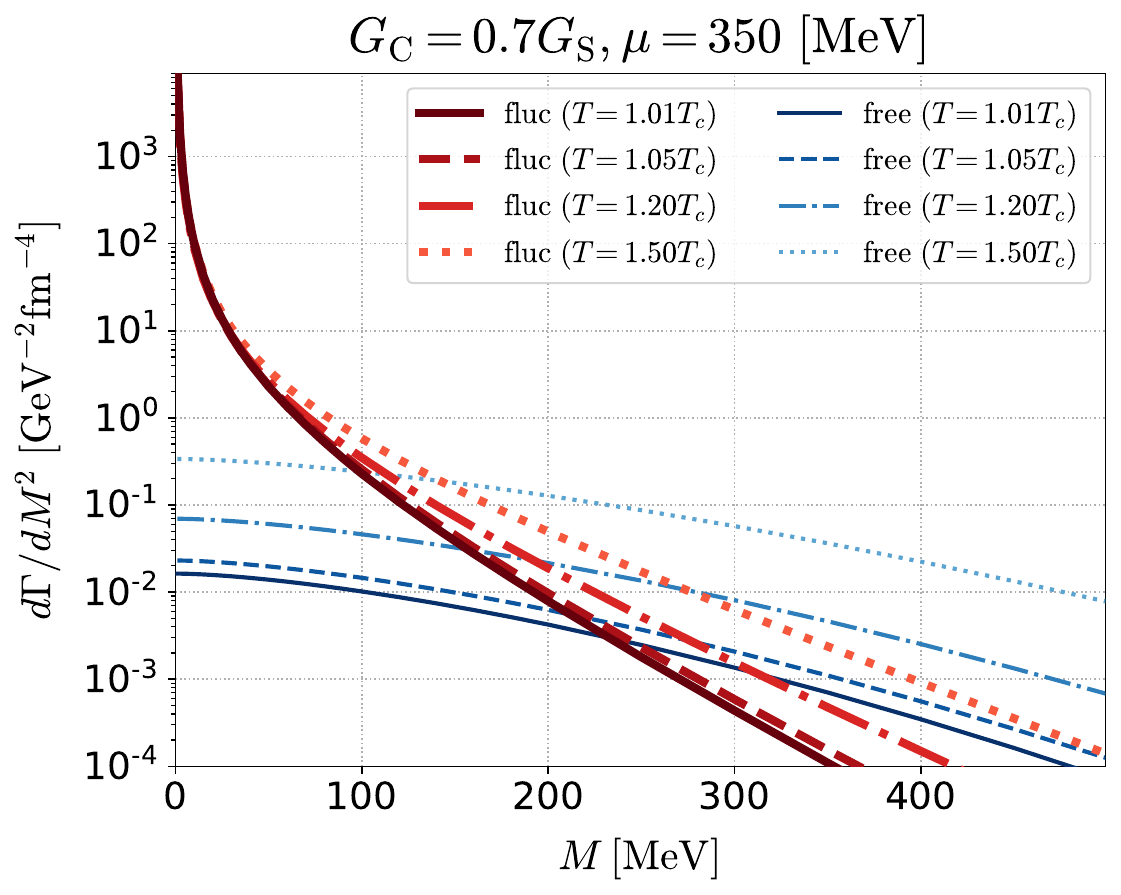}
\includegraphics[width=0.45\linewidth]{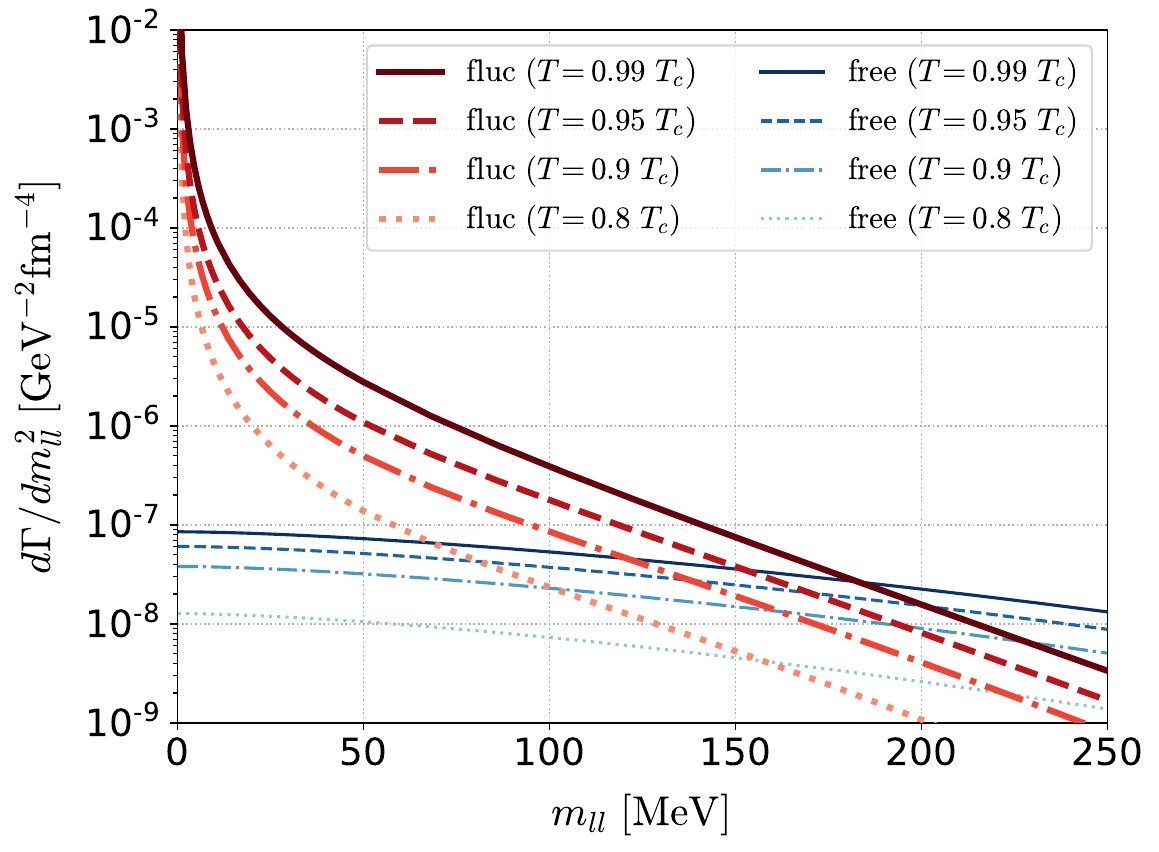}
\caption{
Dilepton production rates per unit energy $\omega$ and momentum $\bm{k}$ above $T_c$ of the 2SC at $\mu = 350~{\rm MeV}$ (left)~\cite{Nishimura:2022mku} and of the QCD-CP at $\mu = \mu_{\rm CP}$ (right)~\cite{Nishimura:2023oqn} with $G_D = 0.7G_S$. 
The thick (thin) lines are the contribution of the soft modes (the massless free quark gases). 
}
\label{fig:DPRM}
\end{figure}

Finally, we focus on the DPR, which is an experimentally observable quantity in the HIC. As the DPR is extracted from the photon self-energy as in Eq.~\eqref{eq:DPR}, we calculate it using $\Pi^{{\rm R}ij}(\bm k,\omega)$ obtained in Sec.~\ref{sec:Pi}.

The left panel of Fig.~\ref{fig:DPR} shows the DPR 
per unit energy $\omega$ and momentum $\bm{k}$
near the 2SC-PT for $T/T_c=1.01$, $1.1$ and $1.5$ at $\mu=350~{\rm MeV}$ 
and $G_D = 0.7G_S$~\cite{Nishimura:2022mku}.
The thick lines are the contributions from the soft modes, 
while the thin lines are those of the free quark gas.
One sees that the DPR from the soft modes is anomalously enhanced 
at small $\omega$ and $\bm{k}$ region in comparison 
with the free quark gas for $T\lesssim1.5T_c$,
and this enhancement is more pronounced as $T$ approaches $T_c$.
This result is expected from the properties of the soft modes.

In HIC, the DPR is observed as a function of the invariant mass $M=\sqrt{\omega^2-\bm{k}^2}$ to remove the effects of the motion of the medium.
Shown in Fig.~\ref{fig:DPRM} is the invariant-mass spectra of DPR near the 2SC-PT (left) and QCD-CP (right)~\cite{Nishimura:2023oqn}.
One finds that the anomalous enhancement of the DPR is observed at the low-mass region $M\lesssim200$~MeV. 

The low energy-momentum limit of the DPR is related to the electric conductivity 
as is verified in Eqs.~\eqref{eq:sigma} and~\eqref{eq:DPR}. 
Therefore, the dilepton production in the HIC is enhanced when the medium 
created by the collisions passes through the red-color region 
in the figure.
The existence of the two hot spots, corresponding to the 2SC-CP and QCD-CP, 
in Fig.~\ref{fig:sigma_3d} may suggest that the beam-energy scan can measure distinct 
peaks of the DPR~\cite{Nishimura:2023not}.

\section{Brief summary and concluding remark}

In this article, 
we have made a unified account of
the soft modes of the QCD critical point (QCD-CP) and the critical point of the 2SC phase transition (2SC-CP) and their relevance to relativistic heavy-ion collisions (HIC),
based on the 2-flavor Nambu-Jona-Lasinio (NJL) 
model. 
We started by discussing not only static but also dynamical fluctuations of 
physical quantities coupled to the order parameters of these second-order 
phase transitions in the normal phase in a comprehensive way, 
and then showed the very existence of the soft modes for  both the transitions
within the mean-field level calculations.
Then it was demonstrated that the 
soft modes affect various observables and cause interesting phenomena near the QCD-CP and 2SC-CP.
Firstly, it was shown that the diquark soft mode of the 2SC 
gives rise to a ``pseudogap'' in the quark density of states around the 
Fermi surface in the normal phase just above
the critical temperature of the 2SC. Although the appearance of the pseudogap is
of great interest in relation to the similar phenomena seen in condensed matter physics,
it is left as a future problem to identify good observables
to confirm it experimentally.
As experimentally feasible observables, we took up
the electromagnetic observables, such as electric conductivity and 
the dilepton production rate in heavy-ion collisions, and showed that
these quantities are largely affected by the soft modes 
 so that they both get to increase in a divergent way when the system
approaches the respective critical temperature 
from the normal phase. For the analyses of these quantities, we have extended the ideas that are successful for 
account for the ``para-conductivity'' in the normal phase of metallic superconductors.

\vspace{.5cm}
\authorcontributions{}

Both authors contributed equally to shaping the project and research plan. 
M.~K made the major contribution to the actual computations and made the figures, 
and wrote the first draft of the main text except for Abstract, Introduction, 
and the last section, which were written by T.~K.
The authors equally contributed to finalizing 
the whole text of the paper.

\acknowledgments{}

A major part of this study is based on the collaboration with Toru Nishimura, for whom the authors acknowledge. 
This work was partially supported in part by JSPS KAKENHI (Nos. JP22K03619, JP24K07049), ISHIZUE 2025 of Kyoto University, and the Center for Gravitational Physics and Quantum Information (CGPQI) at Yukawa Institute for Theoretical Physics.

\conflictsofinterest

The authors declare no conflicts of interest.

\appendixtitles{no} 




\bibliography{reference.bib}




\PublishersNote{}
\end{document}